\documentclass[longauth]{aa}  

\usepackage{graphicx}
\usepackage{txfonts}
\usepackage{xcolor} 
\usepackage{hyperref} 
\usepackage{longtable}
\usepackage{soul}
\usepackage{ulem}
\begin{document} 


   \title{Dark Energy Survey Deep Field photometric redshift performance and training incompleteness assessment}
   \titlerunning{DES: Deep Field photometric redshift}

   \author{L.~Toribio San Cipriano\inst{1} \and
           J.~De~Vicente\inst{1} \and
           I.~Sevilla-Noarbe\inst{1} \and
           W.~G.~Hartley\inst{2} \and
           J.~Myles\inst{3} \and
           A.~Amon\inst{4, 5} \and
           G.~M.~Bernstein\inst{6} \and
           A.~Choi\inst{7} \and
           K.~Eckert\inst{6} \and
           R.~A.~Gruendl\inst{8, 9} \and
           I.~Harrison\inst{10} \and
           E.~Sheldon\inst{11} \and
           B.~Yanny\inst{12} \and
           M.~Aguena\inst{13} \and
S.~S.~Allam\inst{12} \and
O.~Alves\inst{14} \and
D.~Bacon\inst{15} \and
D.~Brooks\inst{16} \and
A.~Campos\inst{17} \and
A.~Carnero~Rosell\inst{18, 13, 19} \and
J.~Carretero\inst{20} \and
F.~J.~Castander\inst{21, 22} \and
C.~Conselice\inst{23, 24} \and
L.~N.~da Costa\inst{13} \and
M.~E.~S.~Pereira\inst{25} \and
T.~M.~Davis\inst{26} \and
S.~Desai\inst{27} \and
H.~T.~Diehl\inst{12} \and
P.~Doel\inst{16} \and
I.~Ferrero\inst{28} \and
J.~Frieman\inst{12,29} \and
J.~Garc\'ia-Bellido\inst{30} \and
E.~Gaztañaga\inst{15, 21, 22} \and
G.~Giannini\inst{20,29} \and
S.~R.~Hinton\inst{26} \and
D.~L.~Hollowood\inst{31} \and
K.~Honscheid\inst{32,33} \and
D.~J.~James\inst{34} \and
K.~Kuehn\inst{35,36} \and
S.~Lee\inst{37} \and
C.~Lidman\inst{38,39} \and
J.~L.~Marshall\inst{40} \and
J. Mena-Fern{\'a}ndez\inst{41} \and
F.~Menanteau\inst{8, 9} \and
R.~Miquel\inst{42,20} \and
A.~Palmese\inst{17} \and
A.~Pieres\inst{13, 43} \and
A.~A.~Plazas~Malag\'on\inst{44, 45} \and
A.~Roodman\inst{44, 45} \and
E.~Sanchez\inst{1} \and
M.~Smith\inst{46} \and
M.~Soares-Santos\inst{14} \and
E.~Suchyta\inst{47} \and
M.~E.~C.~Swanson\inst{8} \and
G.~Tarle\inst{14} \and
M.~Vincenzi\inst{15, 46} \and
N.~Weaverdyck\inst{14, 48} \and
P.~Wiseman\inst{46} \and
           \\ \vspace{0.2cm} (DES Collaboration)}

   \institute{Centro de Investigaciones Energ\'eticas, Medioambientales y Tecnol\'ogicas (CIEMAT), Madrid, Spain \email{laura.toribio@ciemat.es}
\and
Department of Astronomy, University of Geneva, ch. d'\'Ecogia 16, CH-1290 Versoix, Switzerland 
\and
Department of Astrophysical Sciences, Princeton University, Peyton Hall, Princeton, NJ 08544, USA
\and
Institute of Astronomy, University of Cambridge, Madingley Road, Cambridge CB3 0HA, UK
\and
Kavli Institute for Cosmology, University of Cambridge, Madingley Road, Cambridge CB3 0HA, UK
\and
Department of Physics and Astronomy, University of Pennsylvania, Philadelphia, PA 19104, USA
\and
NASA Goddard Space Flight Center, 8800 Greenbelt Rd, Greenbelt, MD 20771, USA
\and
Center for Astrophysical Surveys, National Center for Supercomputing Applications, 1205 West Clark St., Urbana, IL 61801, USA
\and
Department of Astronomy, University of Illinois at Urbana-Champaign, 1002 W. Green Street, Urbana, IL 61801, USA
\and
School of Physics and Astronomy, Cardiff University, CF24 3AA, UK
\and
Brookhaven National Laboratory, Bldg 510, Upton, NY 11973, USA
\and
Fermi National Accelerator Laboratory, P. O. Box 500, Batavia, IL 60510, USA
\and
Laborat\'orio Interinstitucional de e-Astronomia - LIneA, Rua Gal. Jos\'e Cristino 77, Rio de Janeiro, RJ - 20921-400, Brazil
\and
Department of Physics, University of Michigan, Ann Arbor, MI 48109, USA
\and
Institute of Cosmology and Gravitation, University of Portsmouth, Portsmouth, PO1 3FX, UK
\and
Department of Physics \& Astronomy, University College London, Gower Street, London, WC1E 6BT, UK
\and
Department of Physics, Carnegie Mellon University, Pittsburgh, Pennsylvania 15312, USA
\and
Instituto de Astrofisica de Canarias, E-38205 La Laguna, Tenerife, Spain
\and
Universidad de La Laguna, Dpto. Astrofísica, E-38206 La Laguna, Tenerife, Spain
\and
Institut de F\'{\i}sica d'Altes Energies (IFAE), The Barcelona Institute of Science and Technology, Campus UAB, 08193 Bellaterra (Barcelona) Spain
\and
Institut d'Estudis Espacials de Catalunya (IEEC), 08034 Barcelona, Spain
\and
Institute of Space Sciences (ICE, CSIC),  Campus UAB, Carrer de Can Magrans, s/n,  08193 Barcelona, Spain
\and
Jodrell Bank Center for Astrophysics, School of Physics and Astronomy, University of Manchester, Oxford Road, Manchester, M13 9PL, UK
\and
University of Nottingham, School of Physics and Astronomy, Nottingham NG7 2RD, UK
\and
Hamburger Sternwarte, Universit\"{a}t Hamburg, Gojenbergsweg 112, 21029 Hamburg, Germany
\and
School of Mathematics and Physics, University of Queensland,  Brisbane, QLD 4072, Australia
\and
Department of Physics, IIT Hyderabad, Kandi, Telangana 502285, India
\and
Institute of Theoretical Astrophysics, University of Oslo. P.O. Box 1029 Blindern, NO-0315 Oslo, Norway
\and
Kavli Institute for Cosmological Physics, University of Chicago, Chicago, IL 60637, USA
\and
Instituto de Fisica Teorica UAM/CSIC, Universidad Autonoma de Madrid, 28049 Madrid, Spain
\and
Santa Cruz Institute for Particle Physics, Santa Cruz, CA 95064, USA
\and
Center for Cosmology and Astro-Particle Physics, The Ohio State University, Columbus, OH 43210, USA
\and
Department of Physics, The Ohio State University, Columbus, OH 43210, USA
\and
Center for Astrophysics $\vert$ Harvard \& Smithsonian, 60 Garden Street, Cambridge, MA 02138, USA
\and
Australian Astronomical Optics, Macquarie University, North Ryde, NSW 2113, Australia
\and
Lowell Observatory, 1400 Mars Hill Rd, Flagstaff, AZ 86001, USA
\and
Jet Propulsion Laboratory, California Institute of Technology, 4800 Oak Grove Dr., Pasadena, CA 91109, USA
\and
Centre for Gravitational Astrophysics, College of Science, The Australian National University, ACT 2601, Australia
\and
The Research School of Astronomy and Astrophysics, Australian National University, ACT 2601, Australia
\and
George P. and Cynthia Woods Mitchell Institute for Fundamental Physics and Astronomy, and Department of Physics and Astronomy, Texas A\&M University, College Station, TX 77843,  USA
\and
LPSC Grenoble - 53, Avenue des Martyrs 38026 Grenoble, France
\and
Instituci\'o Catalana de Recerca i Estudis Avan\c{c}ats, E-08010 Barcelona, Spain
\and
Observat\'orio Nacional, Rua Gal. Jos\'e Cristino 77, Rio de Janeiro, RJ - 20921-400, Brazil
\and
Kavli Institute for Particle Astrophysics \& Cosmology, P. O. Box 2450, Stanford University, Stanford, CA 94305, USA
\and
SLAC National Accelerator Laboratory, Menlo Park, CA 94025, USA
\and
School of Physics and Astronomy, University of Southampton,  Southampton, SO17 1BJ, UK
\and
Computer Science and Mathematics Division, Oak Ridge National Laboratory, Oak Ridge, TN 37831
\and
Lawrence Berkeley National Laboratory, 1 Cyclotron Road, Berkeley, CA 94720, USA
}

   \date{Received 14 December 2023; accepted 24 February 2024}

 
  \abstract
   {The determination of accurate photometric redshifts (photo-zs) in large imaging galaxy surveys is key for cosmological studies. One of the most common approaches are machine learning techniques. These methods require a spectroscopic or reference sample to train the algorithms. Attention has to be paid to the quality and properties of these samples since they are key factors in the estimation of reliable photo-zs.}
   {The goal of this work is to calculate the photo-zs for the Y3 DES Deep Fields catalogue using the Directional Neighborhood Fitting (DNF) machine learning algorithm. Moreover, we want to develop techniques to assess the incompleteness of the training sample and metrics to study how incompleteness affects the quality of photometric redshifts. Finally, we are interested in comparing the performance obtained with respect to the EAzY template fitting approach on Y3 DES Deep Fields catalogue.}
   {We have emulated --at brighter magnitude-- the training incompleteness with a spectroscopic sample whose redshifts are known to have a measurable view of the problem. We have used a principal component analysis to graphically assess incompleteness and to relate it with the performance parameters provided by DNF. Finally, we have applied the results about the incompleteness to the photo-z computation on Y3 DES Deep Fields with DNF and estimated its performance.
   }
   {The photo-zs for the galaxies on DES Deep Fields have been computed with the DNF algorithm and added to the Y3 DES Deep Fields catalogue. They are available at \href{https://des.ncsa.illinois.edu/releases/y3a2/Y3deepfields}{https://des.ncsa.illinois.edu/releases/y3a2/Y3deepfields}. Some techniques have been developed to evaluate the performance in the absence of "true" redshift and to assess completeness. We have studied the tradeoff on the training sample between highest spectroscopic redshift quality vs. completeness. Some advantages have been found on relaxing the highest quality spectroscopic redshift requirements at fainter magnitudes in favour of completeness. The result achieved by DNF on the Y3 Deep Fields are competitive with the ones provided by EAzY showing notable stability at high redshifts. It should be noted the good results obtained by DNF for the estimation of photo-zs in deep field catalogues and its suitability for the future LSST and Euclid data, which will have similar depths to the Y3 DES Deep Fields.}
  {}
   \keywords{dark energy -- 
                Galaxies: distances and redshifts --
                Machine Learning
               }

   \maketitle
%

\section{Introduction} 

The arrival of large photometric galaxy surveys such as Sloan Digital Sky Survey \citep[SDSS,][]{2000AJ....120.1579Y}, Dark Energy Survey \citep[DES,][]{2015AJ....150..150F}, Physics of the Accelerating Universe \citep[PAU,][]{2012SPIE.8446E..6DC} or the future projects such as Vera C. Rubin Observatory Legacy Survey of Space and Time \citep[LSST,][]{2009arXiv0912.0201L}, and Euclid \citep{2020A&A...644A..31E}, capable of collecting huge amounts of data, are providing invaluable insight about the Universe. One of the crucial elements for cosmological and astrophysical studies is the estimation of accurate redshifts from photometric information which are essential for many cosmological probes as baryon acoustic oscillation (BAO), weak lensing or galaxy clustering. Spectroscopic surveys --measuring the difference in the wavelength of some spectral lines with respect to their wavelength at rest frame-- provide high precision redshifts, but obtaining spectroscopic redshifts of large samples of astronomical objects is very expensive in terms of observing time. Currently, the Dark Energy Spectroscopic Instrument (DESI) project \citep{2016arXiv161100036D} is capable of measuring thousands of galaxy spectra every night, reducing telescope time. Despite this great advantage long exposure times are still required to obtain good signal-to-noise spectra of faint objects, and photometric data for target selection. An alternative is to measure the fluxes of galaxies with a set of broadband or narrowband filters within an image survey, that is using photometric techniques. These measurements allow us to compute the photometric redshift (also called photo-zs) of a large number of galaxies per image reducing the telescope time at the cost of a lower precision. 

The two main approaches to determining photo-zs are \textit{template fitting} and  \textit{machine learning} methods. Template methods compare the spectral energy distribution (SED) of each galaxy with that of a set of redshifted rest-frame templates looking for the best match \citep[e.g.][]{1999MNRAS.310..540A, 2000ApJ...536..571B, 2000A&A...363..476B, 2006A&A...457..841I}. Machine learning approaches use reference or training galaxy samples whose spectroscopic redshifts are known in order to learn the relationship between magnitudes, colours and redshifts. With this information, machine learning methods can predict the photometric redshift of a set of target galaxies  \citep[e.g.][]{2004PASP..116..345C, 2016PASP..128j4502S, 2013MNRAS.432.1483C, 2016MNRAS.459.3078D}. Neither method is free of difficulties. Template methods depend on synthetic models and the completeness of the template library used in the fitting, while machine learning methods depend on the quality and variety of the training samples. Specifically, the selection of this spectroscopic training sample is one of the most important decisions to obtain accurate photometric redshift estimations in the machine learning approach. Ideally, the spectroscopic sample should be representative of the whole target galaxy sample, covering the same colour-magnitude space. Unfortunately, the galaxy samples whose photometric redshift is to be determined typically include galaxies with deeper magnitudes that are not included in the spectroscopic sample. \cite{2020MNRAS.496.4769H} studied the impact of using incomplete spectroscopic samples in the redshift distribution using the \cite{2008MNRAS.390..118L} algorithm. They show that an incomplete spectroscopic training sample could bias the galaxy redshifts. Moreover, the studies of \citet{2010A&A...523A..31H}, \citet{2017MNRAS.468.4323B}, \citet{2014MNRAS.445.1482S}, \citet{2020MNRAS.499.1587S}, \citet{2016PhRvD..94d2005B} and \citet{2011MNRAS.417.1891A} and \citet{2021FrASS...8...70B} compare different methods for photo-zs estimation. These works suggest that machine learning methods provide more accurate values of photo-zs than template methods as long as there is a sufficiently adequate sample for training. Outside the magnitude and colour space, template methods seem to perform better than machine learning methods because they can generate synthesized spectra without redshift constraint. Everything seems to indicate that the combination of both template and machine learning is the best option to obtain the best photo-zs accuracy of a sample \citep[e.g.][]{2018PASJ...70S...9T, 2019NatAs...3..212S}. 

In this work, we study how the incompleteness in the spectroscopic training sample affects Directional Neighborhood Fitting (DNF) photo-zs algorithm \citep{2016MNRAS.459.3078D} photo-zs, as estimated in the Dark Energy Survey (DES) Year 3 Deep Field sample. DNF is a nearest-neighbour approach for photometric redshift estimation that has become a reference within DES collaboration and included between the five methods to be prioritized in Vera Rubin observatory. To assess the effects of incompleteness, we first derive the relevant parameters to characterize incompleteness, demonstrating how these parameters affect photo-z performance. Then, we show how DNF accounts for incompleteness in the photo-zs errors provided. Finally, we study the incompleteness of the training sample in Y3 DES Deep Fields and compare our results with those obtained from the EAzY template method \citep{2008ApJ...686.1503B}. 

The rest of the paper is organised as follows. In Sect. \ref{sec:samples}, we describe the sample selection and in Sect. \ref{sec:metrics} the metrics used and the description of DNF algorithm. We carried out an analysis of the effects of incomplete training samples on the estimation of photometric redshift in Sect. \ref{sec:ref_effect_incompleteness}. In Sect. \ref{sec:photoz_deepFields}, we estimate photometric redshift for Y3 DES Deep Fields with different training samples. We compare the photo-zs determined by DNF and EAzY in Sect. \ref{sec:DNFvsEAzY}. Finally, we enumerate the conclusions of this work in Sect. \ref{sec:conclusion}.


\section{Data}\label{sec:samples}
\subsection{Spectroscopic sample} \label{sec:spect_sample}
We used the spectroscopic sample defined by \citet{2018AC....25...58G}. This sample contains spectroscopic redshifts of galaxies from 34 surveys (see Appendix \ref{sec:appendixA}) and the photometric information for each of them. The quality of the spectroscopic redshift is flagged by the label $\mathrm{FLAG\_DES}$ (with $\mathrm{FLAG\_DES} = 4$ as certain redshift, $\mathrm{FLAG\_DES} = 3$ as probable redshift, $\mathrm{FLAG\_DES} = 2$ as possible redshift and $\mathrm{FLAG\_DES} = 1$ as unknown redshift). For this work, we only select those objects with spectroscopic redshift determination marked in the catalogue with the best redshift determination, that is: those galaxies with flag of 3 and 4 level ($\mathrm{FLAG\_DES} \geq 3$). In addition, we exclude those galaxies with $mag (i) \geq 28$. After these cuts, our spectroscopic sample contains a total of 55,601 galaxies.

\subsection{Y3 Deep Fields catalogue}

The Y3 DES Deep Fields catalogue\footnote{Available at https://des.ncsa.illinois.edu/releases/y3a2/Y3deepfields} used is part of the Dark Energy Survey. The observations were taken using the Dark Energy Camera \citep[DECam,][]{2015AJ....150..150F} on the Victor M. Blanco 4m telescope at the Cerro Tololo Inter-American Observatory (CTIO) in Chile. DES covered $5000\ \mathrm{deg}^2$ in \textit{$grizY$} bands with approximately 10 overlapping dithered exposures in each filter (90 sec in \textit{$griz$}, 45 sec in \textit{$Y$}) covering the survey footprint. The Y3 DES Deep Fields catalogue comprises four fields measured with 8 bands (\textit{$ugrizJHK_s$}) covering an area of $\sim 5.88\ \mathrm{deg}^2$ where the integrated exposure time per pixel is approximately ten times more than the main DES survey area  \citep[see details in][]{2022MNRAS.509.3547H}. This catalogue contains around 2.8 million galaxies. We have selected those galaxies that have flux measurements in the eight filters and with $mag(i)<28$, resulting a catalogue that contains around 1.5 million galaxies. We have selected galaxies with $mag(i)<28$ --still suitable for weak lensing applications-- because for higher magnitudes the errors in the photometry are large and the data become unreliable.

\section{Metrics and algorithm}\label{sec:metrics}
\subsection{Metrics}
This section describes the metrics used in this work to assess the quality of the photo-zs estimates, where  $z_{spec}$, $z_{phot}$ and $N$ represent the spectroscopic redshift, the photometric redshift and the number of objects in the sample, respectively. We define the following metrics to quantify the degree of precision of the photo-zs and its scatter:

\begin{itemize}

\item{\emph{Bias}}: The assessment of the overall photo-zs is determined by the \emph{mean bias}:

$$ \overline{\Delta z} = \frac{1}{N}\sum_{i=1}^N (\Delta z_i),$$

where $\Delta z= z_{spec} - z_{phot}$.

\item \emph{Mean Absolute Deviation}:

$$\mathrm{MAD} (\Delta z) = \frac{1}{N}\sum_{i=1}^N|\Delta z_i|.$$



\item $\sigma_{68}(\Delta z)$: denotes the half width of the central 68\% percentile range of galaxies both bias value.

$$\sigma_{68}(\Delta z) = \frac{1}{2}(P_{84} - P_{16}),$$ 
where 
$P_{16}=$ 16th  percentile  of  the  cumulative distribution and
$P_{85}=$ 84th percentile  of  the  cumulative  distribution.

\item $\sigma_{68}$ \emph{normalised}: defined as
$$\sigma_{68}^{Norm}= \sigma_{68} \Big(\frac{|\Delta z|}{1 + z_{spec}} \Big).$$

\item \emph{Outlier fraction}: 
$$f = \frac{N_{out}}{N},$$

where $N$ is the total number of objects and $N_{out}$ the outlier defined by: 

$$|\Delta z| \ge 3\sigma, $$

where $\sigma$ is the standard deviation of the $\Delta z$ distribution. 

\end{itemize}

\subsection{DNF algorithm}
Directional Neighborhood Fitting  \citep[DNF,][]{2016MNRAS.459.3078D} is a nearest neighbour algorithm for estimating the redshift of a sample of galaxies. DNF uses the colours and magnitudes/fluxes as a measurement of closeness to a reference sample composed of galaxies whose spectroscopic redshifts are known. DNF provides the main photo-z value and its error estimation along with a secondary value intended for photo-z distribution estimation:
\begin{itemize}
\item $\mathrm{DNF\_Z}$: the main photo-z estimate determined by the fit of a number of neighbour galaxies to a  hyperplane in magnitude space. The process is iterated to remove outliers. In addition the algorithm can provide individual photo-z probability density funcions (PDFs).
\item $\mathrm{DNF\_ZSIGMA}$: indicator of photo-z quality computed from the quadratic sum of the error due to photometry plus the error due to the fit. $\mathrm{DNF\_ZSIGMA}$ takes the value -99 when DNF does not estimate the photo-z of a galaxy because there is no neighbour galaxy within a given radius. 

\item $\mathrm{DNF\_ZN}$: a secondary photo-z determined by the single nearest neighbour galaxy which is valuable for redshift distribution estimation.
\end{itemize}

The algorithm provides three alternative metrics for the assessment of closeness: Euclidean, angular and directional. While Euclidean and angular metrics account for magnitude and colour respectively, directional metric integrate both in a unique number. The present work takes advantage of the combination of 5 optical plus 3 near-infrared filters to define non-degenerated colours within the angular metric.

\section{Effect of training incompleteness in photometric redshift estimation}\label{sec:ref_effect_incompleteness}

\begin{figure}
  \centering
	\includegraphics[width=0.50\textwidth]{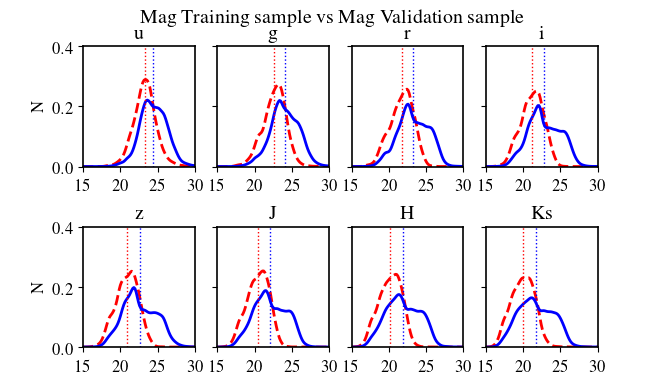}
	\includegraphics[width=0.50\textwidth]{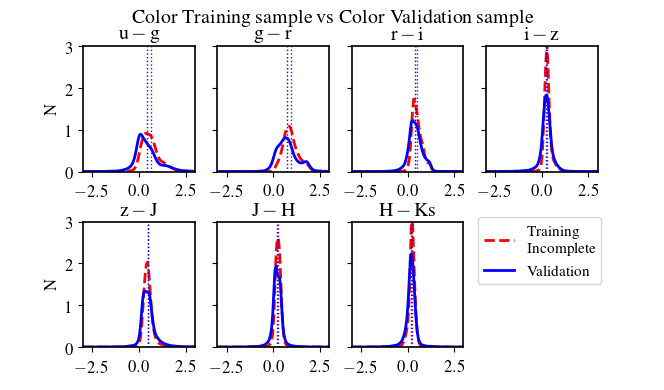}
    \caption{Magnitude and colour distribution for incomplete training and validation samples. The red dashed lines represent the incomplete training sample and the blue lines the validation sample. The dotted vertical lines are the mean of each distribution. We have not included the curve for the complete training figure because these distributions overlap perfectly with those of the validation sample.}
    \label{fig:mag_and_color_distribution}
\end{figure}

We study the effect of using an incomplete spectroscopic training sample to determine the photo-zs with the DNF algorithm. We refer to an incomplete training sample when it does not cover the same range of magnitudes and/or colours as the target sample for which we want to determine its photo-zs. 

The spectroscopic sample, in addition to being used to train the algorithm, allows us to study the accuracy and precision of the photo-z estimation. For this purpose, the spectroscopic sample is usually split into two samples: one used to train the algorithm and the other one to validate the photo-zs (known as the training and validation sample, respectively). However, we must be careful when extrapolating the results obtained in the validation sample to the galaxies in the scientific target sample. The scientific sample may well contain galaxies at deeper magnitudes or in a different colour range which are not represented in the training sample and photo-zs may not be correctly estimated.  

\subsection{Incompleteness emulation with the spectroscopic sample}
\label{sec:emulation}

With the goal to learn how the incompleteness affects photometric redshift performance, we use the spectroscopic sample to emulate, at brighter magnitudes, two different scenarios: a case in which we have completeness of magnitude and colour coverage from the training to the target sample, and another in which we do not (incomplete case).

We split the spectroscopic catalogue into two sub-samples of equal size (with $27,801$ galaxies each) and equal magnitude-colour distribution. We take one of these sub-samples as a validation sample and the other as training sample. We select the galaxies of the training sample in two different ways  to emulate the scenarios mentioned. On one hand, we take all galaxies of the training sample (the $27,801$ galaxies) to emulate a complete training set, that is a training sample that covers the same colour-magnitude space as the validation sample. On the other hand, we use the training sample to construct an incomplete version. In this second case, we want to emulate, at brighter magnitudes, the incompleteness observed in the Y3 DES Deep Fields catalogue when the training sample is formed by galaxies of the spectroscopic sample with FLAG\_DES$=$4. To achieve this, some high-magnitude galaxies can be manually removed from the training sample until incompleteness is reached. In order to automate this process rather than performing it manually, we have employed the following method. We first calculate $\Delta_{band}$, that is the difference between the mean magnitude of the objects in the spectroscopic sample and in the Y3 DES Deep Fields photometric catalogue, for each band. Then, we subtract $\Delta_{band}$ from the magnitudes of every galaxy in the training sample to emulate a similar incompleteness at brighter magnitudes. Applying this magnitude left-shift, we achieve a magnitude incompleteness at the expense of decoupling galaxies from their own redshift. To solve this issue, we use a nearest neighbour algorithm to find, within the shifted sample, real galaxies with similar magnitudes. The algorithm assigns for each left-shifted magnitude a real galaxy from the training sample, many of them repeated. After applying this procedure and dropping out the repeated galaxies, the new training sample, hereafter referred to as the incomplete training sample, is reduced to $5,336$ galaxies out of the original $27,801$. In this way, we now have a galaxy sample that simulates incompleteness in a magnitude range for which we have information about the spectroscopic redshift, enabling us for the study of the effects of incompleteness.

\begin{figure}
  \centering
	\includegraphics[width=0.50\textwidth]{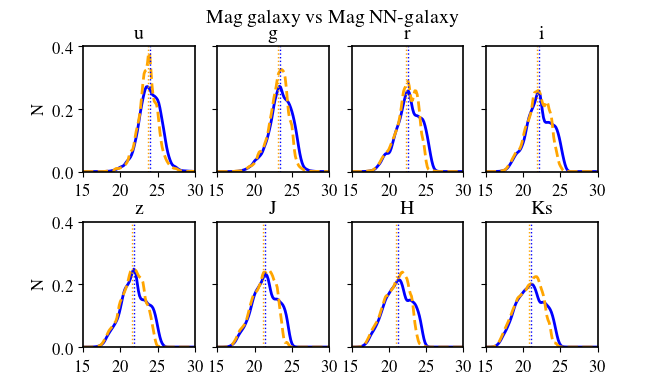}
	\includegraphics[width=0.50\textwidth]{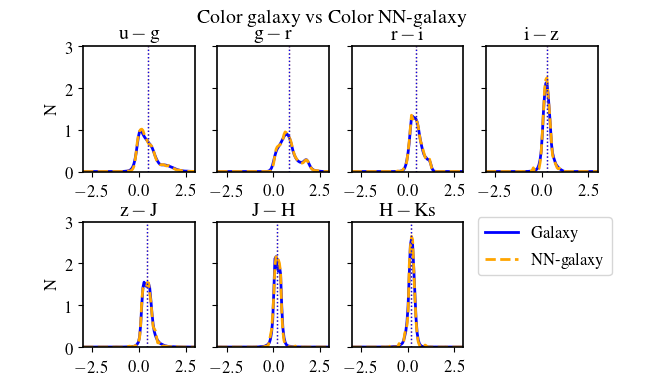}
    \caption{Magnitude and colour distribution for the nearest neighbour galaxies used for the estimation of photo-z in the incomplete training and the validation sample. The orange dashed lines represent the nearest neighbour galaxies distribution of incomplete training and the blue lines the validation sample. The dotted vertical lines are the mean of each distribution. We have not included the distributions for the complete training sample since they overlap perfectly with those of the validation sample.}
    \label{fig:mag_and_color_distribution_NN}
\end{figure}

Figure~\ref{fig:mag_and_color_distribution} shows the magnitudes and colour distributions (upper and lower plots, respectively) for the incomplete training sample (red dashed lines) and the validation sample (blue lines). We have not included the comparison to the complete training sample since their distributions overlap perfectly with those of the validation sample by construction.

\subsection{Incompleteness assessment}\label{sec:incomplete_ass}

\begin{figure}
  \centering	\includegraphics[width=0.49\textwidth]{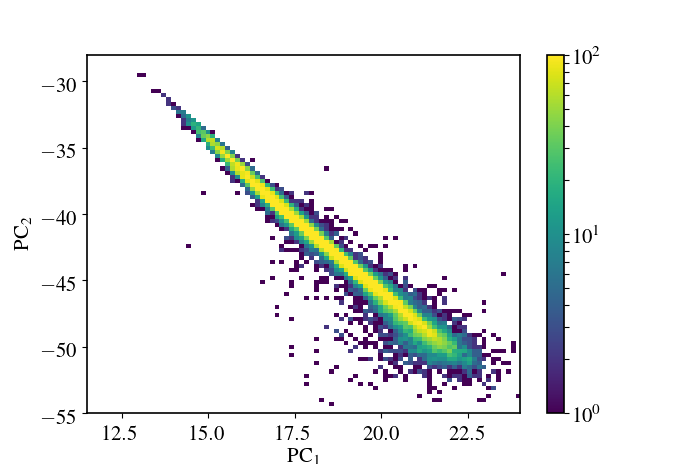}
	\includegraphics[width=0.49\textwidth]{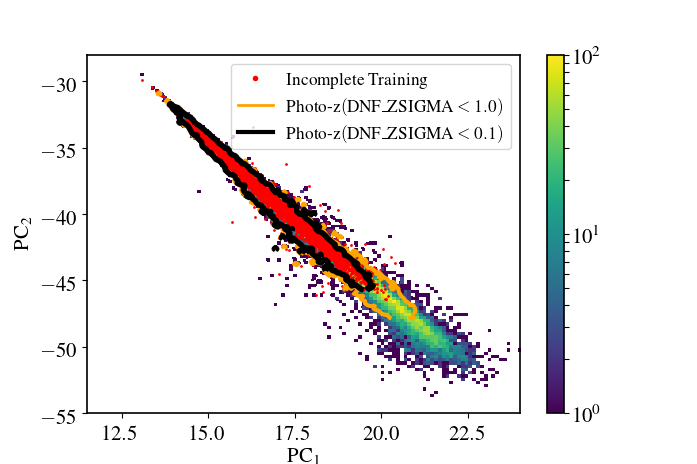}
    \caption{Density map as a function of the first and second principal component of the galaxies of the validation sample. In the bottom panel, we included the galaxies of the training sample (red dots) and the limit in principal components of the galaxies that DNF provides a value of photoz with $\mathrm{DNF\_ZSIGMA <1.0}$ (orange line) and $\mathrm{DNF\_ZSIGMA <0.1}$ (bold black line) with this training sample.
    }
    \label{fig:principal_components}
\end{figure}

\begin{figure*}
  \centering
  	\includegraphics[width=0.49\textwidth]{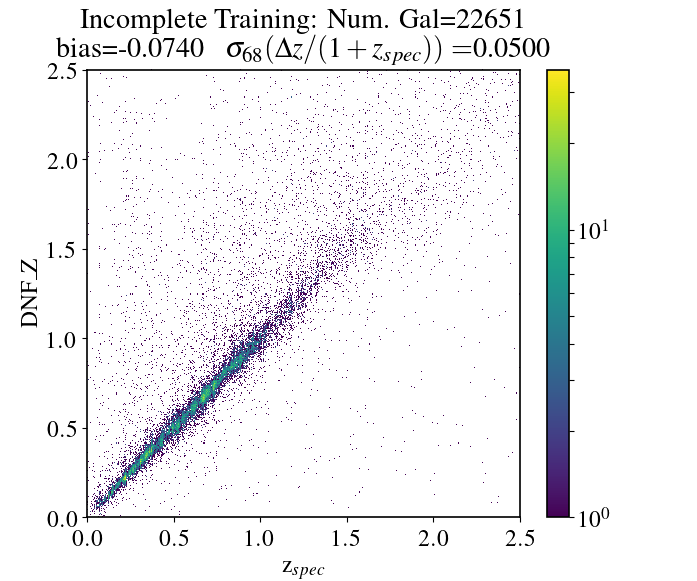}
	\includegraphics[width=0.49\textwidth]{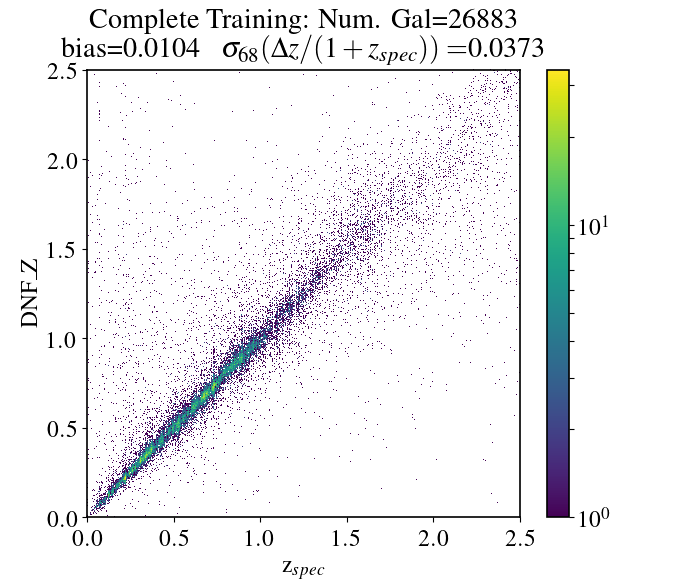}
    \caption{Scatter plot of spectroscopic redshift $z_{spec}$ and the photo-z $\mathrm{DNF\_Z}$ for incomplete training (left panel) and complete training (right panel). Note the improvement for $z>1$ in the complete case.}
    \label{fig:scatter}
\end{figure*}

We determine the DNF photometric redshifts for the validation sample with both complete and incomplete training sets defined as in Sect. \ref{sec:emulation}. We select objects meeting the conditions $\mathrm{DNF\_Z}>0$, $\mathrm{DNF\_ZN}>0$ and $\mathrm{DNF\_ZSIGMA}<1.0$ to ensure the quality of the sample. The cut-off of $\mathrm{DNF\_ZSIGMA}$ has been defined taking into account the analysis carried out in Appendix \ref{sec:appendixB} which studies the possible biases that DNF\_ZSIGMA may have as a quality estimator of DNF photo-z. The number of galaxies after these cuts is $26,882$ galaxies ($96.7$\% of the sample) using the complete training sample and $22,617$ ($81.3$\% of the sample) using the incomplete training sample. 

Figure~\ref{fig:mag_and_color_distribution_NN} shows the magnitude and colour distributions of the galaxies in the validation sample (blue lines) versus the distributions of their nearest-neighbours galaxies (orange dashed lines) determined from the incomplete training sample. Note that while nearest-neighbour magnitude distributions do not match to the weaker magnitudes in all the filters, the colour distributions are close to being recovered in comparison with those shown in Fig.~\ref{fig:mag_and_color_distribution}. The matching of the colour distributions between the validation sample and their nearest-neighbour in the incomplete training sample may be a necessary condition to produce a reliable photometric redshift distribution. However, it may not be sufficient due to the possibility of the existence of galaxies with colour combinations not covered by the training sample.

\begin{figure}
   \centering
   \includegraphics[width=0.49\textwidth]{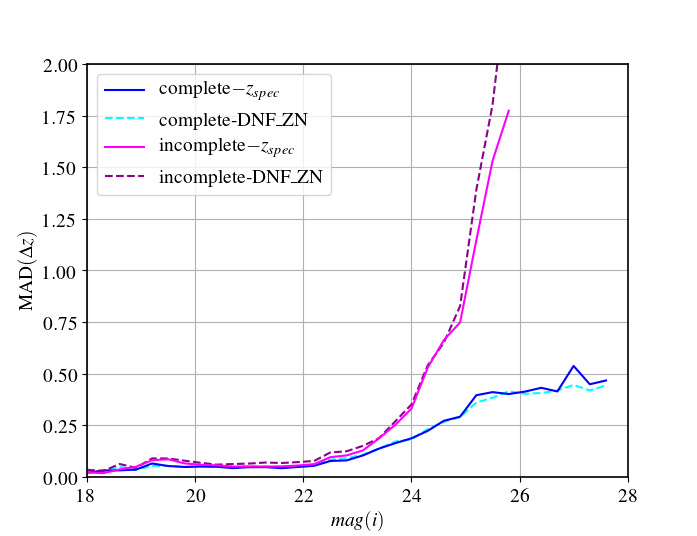}
  \includegraphics[width=0.49\textwidth]{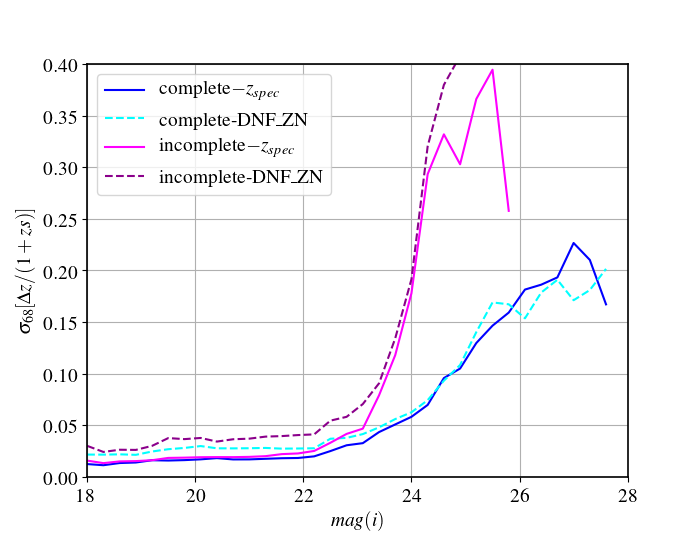}
   \caption{The precision of the photo-z defined by $\mathrm{MAD}(\Delta z)$ (upper panel) and the $\sigma_{68}^{Norm}$ (bottom panel) as a function of the $mag(i)$ for the complete training sample (blue lines) and incomplete training sample (magenta lines). The solid lines display the metrics calculated with $z_{spec}$ and the dashed lines has been calculated replacing $z_{spec}$ by DNF\_ZN.}
   \label{fig:mag_deltaz}%
\end{figure}

In order to study the effect of incompleteness and how to detect it, we have carried out a principal component analysis (PCA). The PCA has been performed with the magnitudes of the bands \textit{$ugrizJHK_s$}. The first principal component ($\mathrm{PC_1}$) for this sample represents $92.8\%$ of the variance of the validation sample in magnitude space, while the percentage grows up to $98.1\%$ with the second component ($\mathrm{PC_2}$). We represent the density map of the validation sample for the principal components in the upper panel of Fig. \ref{fig:principal_components}. We have also stored the first two eigenvectors obtained for  the validation sample to represent in the same basis the training sample. In the bottom panel of Fig.~\ref{fig:principal_components} the red dots show the scatter of the incomplete training sample represented using the same eigenvectors of the validation sample. Comparing both panels of Fig.~\ref{fig:principal_components}, it can be seen that the incomplete training sample does not cover the full validation sample, but this red area is well delimited by the inner bold black line that corresponds to the region for which $\mathrm{DNF\_ZSIGMA <0.1}$. Note that this plot shows the limitations in determining the photo-zs using the DNF algorithm with an incomplete training sample but it also shows how DNF\_ZSIGMA informs of this fact. The outer orange line corresponds to galaxies with $\mathrm{DNF\_ZSIGMA <1.0}$ (this is $81.3$\% of the sample). Therefore, we can identify 3 groups of galaxies. Those galaxies covered by the red dots will have precise photo-zs since for these galaxies the training sample covers the full range of principal component. On other hand, DNF tags as unreliable photo-zs those galaxies outside the orange limit. Therefore, we must study the quality of the photo-zs of the galaxies that are located inside the orange limit but are not covered by the training sample (red area).

\subsection{Photo-zs performance estimation}\label{sec:photo-z_estimation_perf}

\begin{table}
      \caption[]{Summary of metrics.}
         \label{tab:metrics}
$$
         \begin{array}{lccc}         
            \hline
            \noalign{\smallskip}
            \mathrm{Training}\ \mathrm{Sample}  &  \mathrm{Num.}\ \mathrm{galaxies} &  \overline{\Delta z} & \sigma_{68}^{Norm}\\
            \noalign{\smallskip}
            \hline
            \noalign{\smallskip}
            \mathrm{Complete} & 26,883 & 0.0104 & 0.0373  \\
            \mathrm{Incomplete} & 22,651 & -0.0740 & 0.0500  \\
            \noalign{\smallskip}
            \hline
         \end{array}
$$
\end{table}

We compare the photo-z estimation given by DNF in the validation sample using the incomplete and complete training samples. Figure~\ref{fig:scatter} shows the comparison between the spectroscopic redshift ($z_{spec}$) and the photometric redshift ($\mathrm{DNF\_Z}$) for incomplete training sample (left panel) and complete training sample (right panel). We can see that the complete training sample not only determines photo-z values for a larger number of galaxies compared to the incomplete case ($26,883$ galaxies versus $22,651$), but  also presents a lower bias and $\sigma_{68}^{Norm}$ (see Table \ref{tab:metrics}). In addition to the completeness, the number of galaxies in the training sample is a factor that influences the quality of the photo-zs. In Appendix \ref{sec:appendixB2} we have included the results of calculating the photometric redshift using a complete sample with the same number of galaxies as the incomplete sample. The results show that the completeness allows to calculate more accurate photo-zs then the incomplete case for comparables training sample sizes.

\begin{figure}
   \centering
   \includegraphics[width=0.49\textwidth]{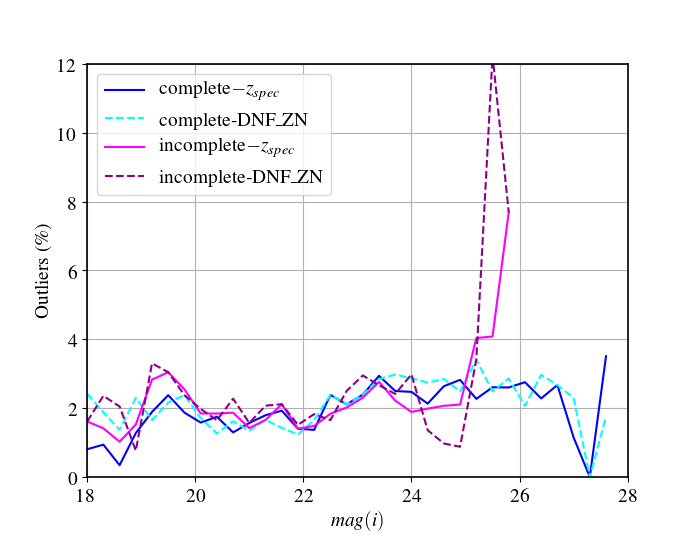}
   \caption{Outliers as a function of the $mag(i)$ for the complete training (blue lines) and incomplete training (magenta lines). The solid lines display the outliers calculated with $z_{spec}$ and the dashed lines replacing $z_{spec}$ by DNF\_ZN.}
   \label{fig:mag_outliers}%
\end{figure}

We have studied the behaviour of the photo-zs estimation through the mean absolute deviation, the $\sigma_{68}^{Norm}$ and outliers. Figure~\ref{fig:mag_deltaz} shows the mean absolute deviation (upper panel) and the $\sigma_{68}^{Norm}$ (bottom panel) of $\mathrm{DNF\_Z}$ with respect to $z_{spec}$ and $\mathrm{DNF\_ZN}$ (solid and dashed lines, respectively) as a function of the $mag(i)$ for the complete training sample (blue lines) and incomplete training sample (magenta lines). The solid lines display the mean absolute deviation and $\sigma_{68}^{Norm}$ of the photo-zs calculated with $z_{spec}$ (which we will refer to as real metric values). It can be readily seen in Fig.~\ref{fig:mag_deltaz} that the completeness of the training sample affects the metrics. Then, we cannot assume that the metrics (mean absolute deviation, $\sigma_{68}^{Norm}$ or those chosen in the study) will have the same behaviour in the validation sample and in the target sample if the training sample shows incompleteness. We must keep in mind that the $z_{spec}$ value of each galaxy is not available when we are calculating the photo-zs for a catalogue so we will not have these measurements to estimate the precision of the photo-zs. Nevertheless, note that $\mathrm{DNF\_ZN}$ (nearest neighbour photo-z) is able to reproduce $z_{spec}$ distribution for moderate training incompleteness in the same way that colour distributions are well recovered for the case of incomplete training (Fig. \ref{fig:mag_and_color_distribution_NN}). In this way, statistical metrics involving $z_{spec}$ are well represented by $\mathrm{DNF\_ZN}$. For this reason, we have calculated the mean absolute deviation and the $\sigma_{68}^{Norm}$ replacing $z_{spec}$ by $\mathrm{DNF\_ZN}$ (dashed lines). In both plots the behaviour of the mean absolute deviation and the $\sigma_{68}^{Norm}$ can be considered a good approximation to the real value which changes depending on the training sample. We can take these metrics calculated by $\mathrm{DNF\_ZN}$ as an upper limit of real ones. Figure~\ref{fig:mag_outliers} shows the outliers as a function of the $ i$ band magnitude $mag(i)$ for the complete training case (blue lines) and incomplete training case (magenta lines). The number of outliers is less than $4\%$ in both cases, up until $mag(i)>24$ where it starts to increase for the incomplete training case. Finally, we complete this study with the behaviour of the photo-z estimation as a function of the spectroscopic redshifts in Appendix \ref{sec:appendixC}.


\section{Photometric redshift Deep Fields catalogue}\label{sec:photoz_deepFields}

\begin{figure*}
   \centering
   \includegraphics[width=0.49\textwidth]{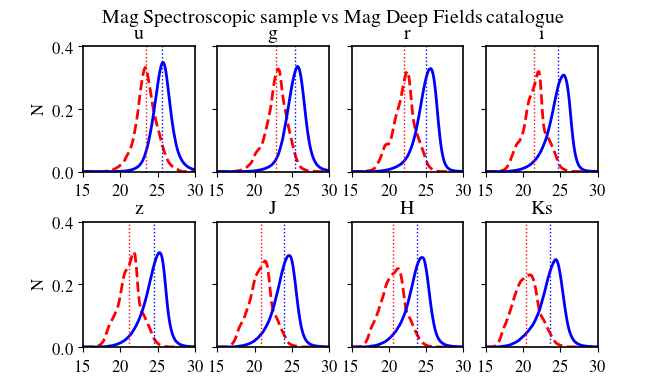} 
   \includegraphics[width=0.49\textwidth]{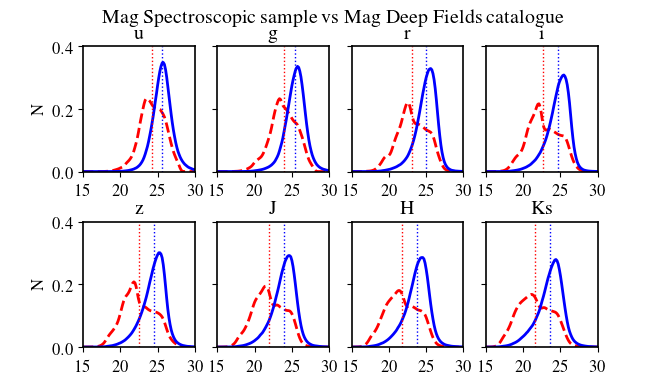}
   \includegraphics[width=0.49\textwidth]{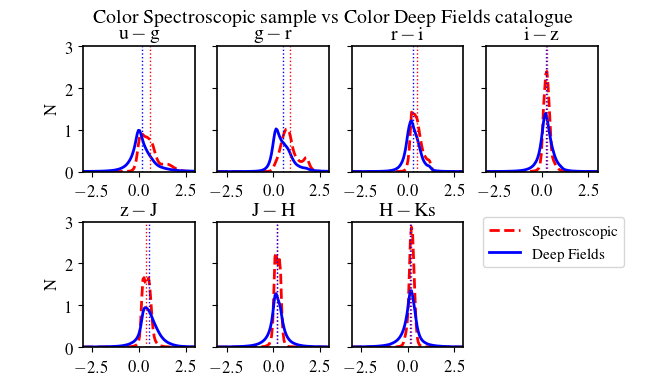}
   \includegraphics[width=0.49\textwidth]{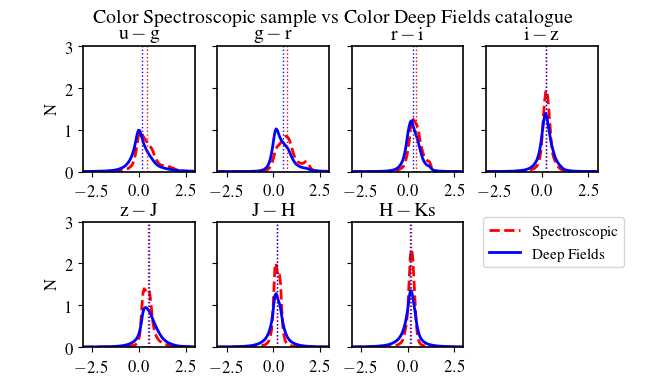}
   \caption{Magnitude and colour distribution for incomplete training (right panels) and semi-complete training (left panels). The red dashed lines represents the training samples and the blue lines the Y3 DES Deep Fields sample. The dotted lines are the mean of each distribution.}
   \label{fig:distribution_mag_df}%
\end{figure*}

We want to study the effects of using different training samples on the quality of the photo-zs in the Y3 DES Deep Fields catalogue. For that, we applied the same analysis developed in Sect. \ref{sec:ref_effect_incompleteness} using two training samples. The first training sample contains only galaxies with the highest quality of spectroscopic redshift determination (i.e. with $\mathrm{FLAG\_DES} = 4$). In this case, the training sample does not contain galaxies with magnitudes as deep as the Y3 DES Deep Fields catalogue. In other words, this training sample is of the highest quality but shows a certain incompleteness with respect to the science sample. The second training sample contains galaxies labelled with spectroscopic redshift quality $\mathrm{FLAG\_DES} \geq 3$. The inclusion  of galaxies whose spectroscopic redshift quality is not optimal but still good in this training sample reduces the problem of incompleteness. In this second case, the training sample reaches the deepest magnitudes of the Y3 DES Deep Fields catalogue. Figure \ref{fig:distribution_mag_df} shows the magnitude and colour distributions of both training samples (in the left panels for the incomplete sample and in the right panels for the semi-complete sample). In order to carry out a similar analysis of that performed in Sect. \ref{sec:ref_effect_incompleteness}, we selected those galaxies with $mag(i)<28.0$ and with a positive flux measurement in the eight filters. This sample contains $1,478,705$ galaxies.

\subsection{Assessment of high quality but incomplete training}\label{sec:ass_complete_train}

For this study, the incomplete training sample is limited to  $38,123$ galaxies for which their spectroscopic redshift has been determined with very high quality. As we can see in the left panels of Fig. \ref{fig:distribution_mag_df}, this spectroscopic sample is shallower than the Y3 DES Deep Fields catalogue (red lines and blue lines, respectively). We want to know how this incompleteness affects the photometric redshift calculation. Using DNF and selecting galaxies with $\mathrm{DNF\_Z}>0$, $\mathrm{DNF\_ZN}>0$ and $\mathrm{DNF\_ZSIGMA}<1.0$, we have determined the photometric redshift for $1,254,981$ galaxies (84.9\%) of the Deep Fields catalogue using this training sample.

\begin{figure*}
   \centering
    \includegraphics[width=0.49\textwidth]{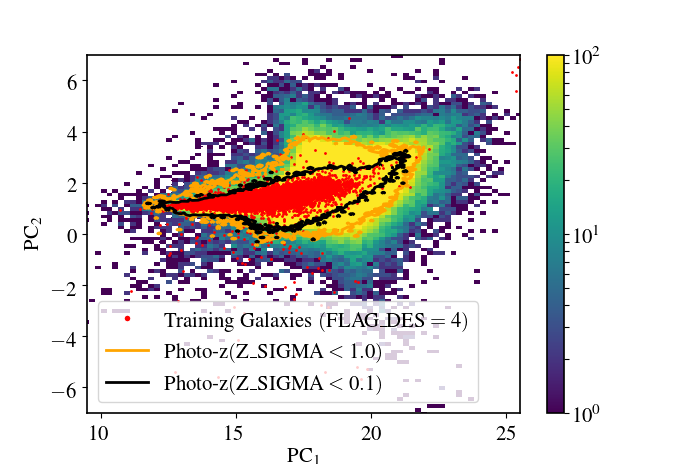} 
    \includegraphics[width=0.49\textwidth]{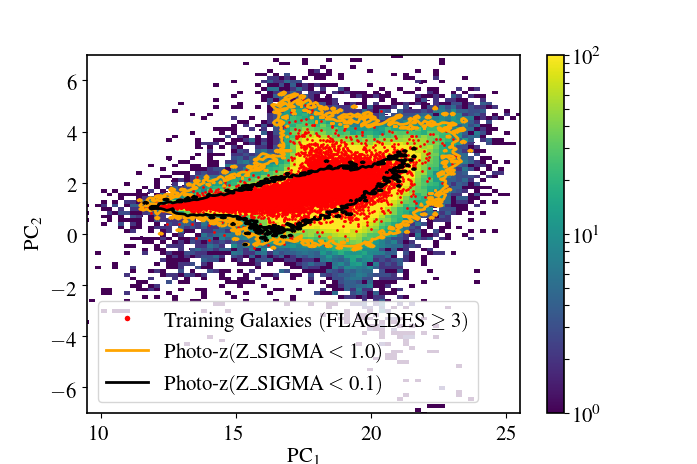} 

   \caption{Density map as a function of the first and second principal component of the galaxies of Y3 DES Deep Fields (density plot in green and yellow colours), the galaxies of the training sample (red dots) and the limit in principal components of the galaxies that DNF provides a value of photoz with $\mathrm{DNF\_ZSIGMA <1.0}$ (orange line) and $\mathrm{DNF\_ZSIGMA <0.1}$ (bold black line) with this training sample. The red blob is less extensive than the green-yellow blob (where the highest density of galaxies is located) when we select only the galaxies with $\mathrm{FLAG\_DES} = 4$.}
   \label{fig:deepfield}%
\end{figure*}

The left panel of Fig.~\ref{fig:deepfield} shows the density map as a function of the first and second principal component of the Y3 DES Deep Fields catalogue and the galaxies of the training sample (red dots). The orange line is the limit of the photo-zs of this 84.9\% of galaxies with the cuts defined above. In addition we overplot another limit bold black line) using a more stringent cut, namely $\mathrm{DNF\_ZSIGMA}<0.1$ that keeps $441,144$ galaxies, that is 29.8\%.

\subsection{Assessment of medium quality but semi-complete training}\label{sec:ass_incomplete_train}
The second training sample used to determine the photometric redshift of Y3 DES Deep Fields catalogue contains $55,601$ galaxies which have magnitudes as deep as the Y3 DES Deep Fields catalogue but with a different distribution as shown in the right panels of Fig.~\ref{fig:distribution_mag_df}.

We can see in the right panel of Fig.~\ref{fig:deepfield}, corresponding to the principal components (the first two eigenvectors represent 93.59\% of the sample), that the spectroscopic training sample is located in the area where the density of galaxies is higher. Although it does not cover the entire of principal component area of the field, DNF provides photo-z for almost all galaxies in the sample ($1,318,960$ galaxies, 91.67\%) delimited by the orange line in the figure. We plot another limit with a bold black line that represents galaxies with a more stringent cut of $\mathrm{DNF\_ZSIGMA}<0.1$ as we did before (405,854 galaxies, i.e. 28.2\%).

\subsection{Performance and comparison of science sets with different training samples}

According to the results obtained, based on the cuts defined in Sect. \ref{sec:ref_effect_incompleteness}, DNF determines photometric redshifts for slightly fewer galaxies when using the incomplete but high quality training sample than in the semi-complete case. The question is how the quality of these photometric redshifts estimates compare. Or in other words, whether it is more important to have high quality spectroscopic redshifts or we can slightly relax that condition to cover the magnitude-colour space as much as possible. 

\begin{figure}
   \centering
    \includegraphics[width=0.49\textwidth]{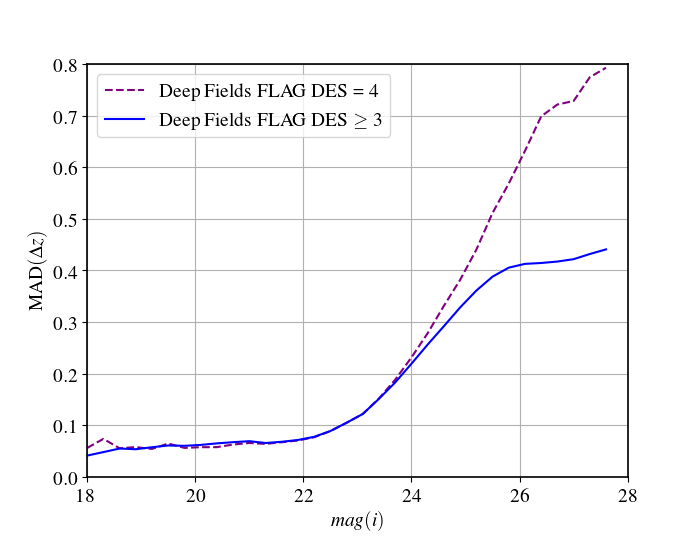} 
    \includegraphics[width=0.49\textwidth]{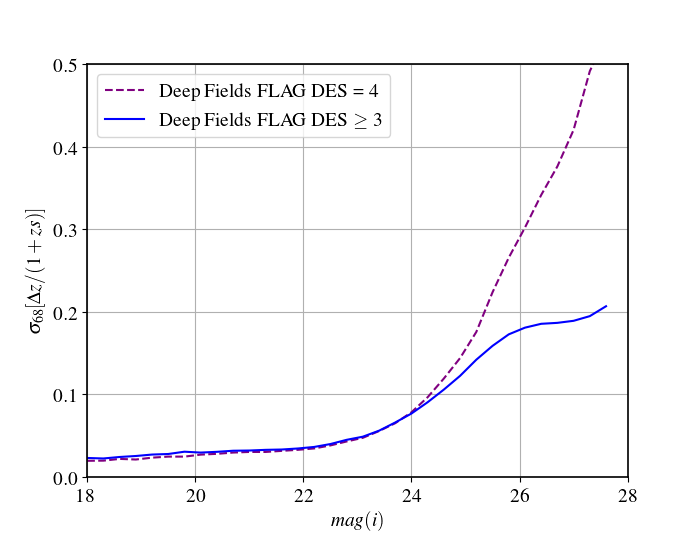}

  \caption{The precision of the photo-z estimates defined by $\mathrm{MAD}(\Delta z)$ (upper panel) and $\sigma_{68}^{Norm}$ (bottom panel) as a function of the $mag(i)$ calculated by $\mathrm{DNF\_Z}$ and $\mathrm{DNF\_ZN}$ determined by training sample with only galaxies with FLAG\_DES $=$ 4 (incomplete training sample, in purple) and training sample with galaxies with FLAG\_DES$\ge$3 (semi-complete training sample, in blue).}
   \label{fig:deepfield_metrics}%
\end{figure}

The results of Fig.~\ref{fig:deepfield_metrics} show the precision of the photo-z estimation by DNF for Y3 DES Deep Fields catalogue defined by mean absolute deviation (left panel) and the $\sigma_{68}^{Norm}$ (right panel) as a function of the $mag(i)$. Note that the $z_{spec}$ of each galaxy is not available in Y3 DES Deep Fields catalogue, then to estimate the mean absolute deviation and the $\sigma_{68}^{Norm}$, we have replaced $z_{spec}$ by $\mathrm{DNF\_ZN}$ following the analysis done in Sect. ~\ref{sec:photo-z_estimation_perf}. We can see that the results obtained by the incomplete, high quality training (dashed purple lines) and the semi-complete training (blue lines) samples follow a similar behaviour for $mag(i) < 24$ although slightly better for the incomplete, high quality training. In this case we obtain a lower error for magnitudes-colour areas covered by the spectroscopic sample. 

We have also seen the same behaviour in Sect. \ref{sec:ass_complete_train} and \ref{sec:ass_incomplete_train} where the incomplete, high quality training contains more galaxies with $\mathrm{DNF\_ZSIGMA <0.1}$ even though, globally, the semi-complete training generates more precise photo-zs. For $mag(i) \ge 24$, the semi-complete training sample, formed by galaxies with slightly lower quality spectroscopic redshift, outperforms the photo-zs of the incomplete training sample formed by the highest quality spectroscopic redshift galaxies. The results indicate that completeness plays an important role in determining higher quality photometric redshift values, as expected. But the results also suggest that for specific studies focussed on brighter galaxies, we may be more interested in using only the redshifts of the highest possible quality in our training. 

Finally, we studied the absolute median deviation and $\sigma_{68}^{Norm}$ as a function of the redshift for the two training samples. You can see more details in Appendix \ref{sec:appendixC}.

\section{Comparison between DNF and EAzY}\label{sec:DNFvsEAzY}

We estimated the photo-zs for the whole deep fields and we added this information to the Y3 DES Deep Fields data\footnote{Available at \href{https://des.ncsa.illinois.edu/releases/y3a2/Y3deepfields}{des.ncsa.illinois.edu/releases/y3a2/Y3deepfields}}. The training sample used to estimate the photo-zs contains galaxies with spectroscopic redshift information labeled with $\mathrm{FLAG\_DES} \geq 3$ corresponding to the semi-complete training sample in Sect.~\ref{sec:ass_incomplete_train}. It is important to note that when computing DNF in this case, we ignore the own spectroscopic redshift for each galaxies in the training sample, to provide a homogeneous comparison of all estimates. 

In addition to the DNF photo-zs \citep{2016MNRAS.459.3078D}, the Y3 DES Deep Fields catalogue contains photo-zs determined with the EAzY algorithm \citep{2022MNRAS.509.3547H, 2008ApJ...686.1503B}. These two methods approach the photometric redshift problem from different perspectives: EAzY determined the photo-zs by fitting a linear combination of template components while DNF is a machine learning code.

\begin{figure*}
   \centering
    \includegraphics[width=0.49\textwidth]{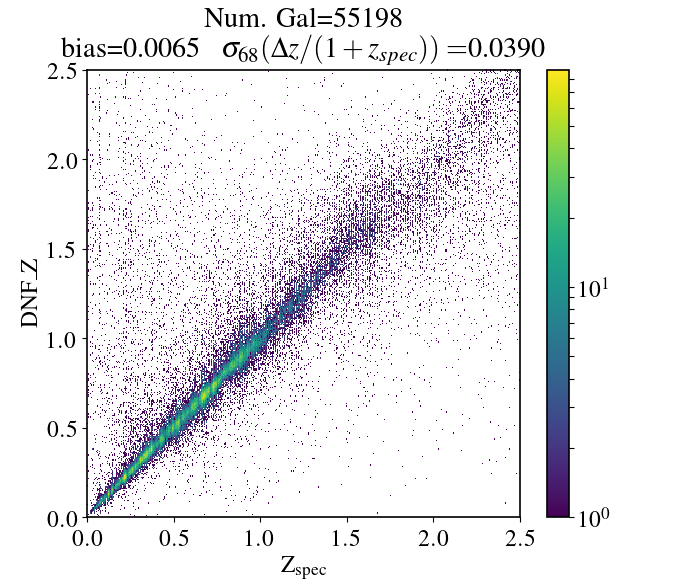} 
   \includegraphics[width=0.49\textwidth]{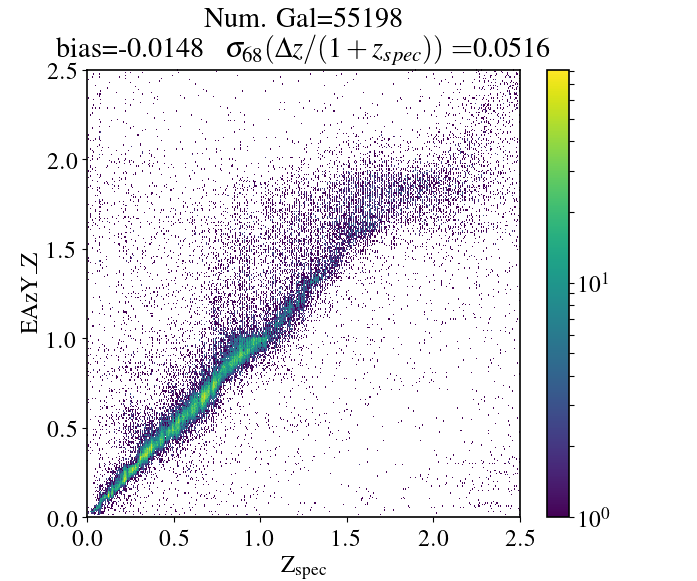}
   \includegraphics[width=0.49\textwidth]{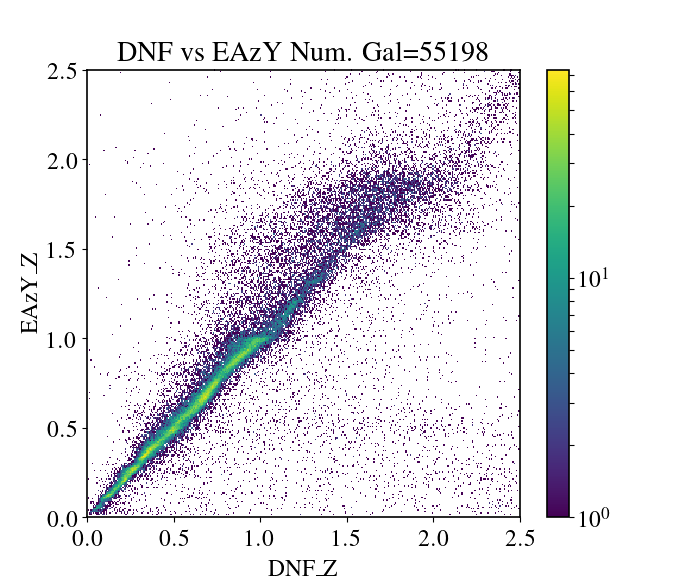} 
   \caption{Scatter plot of spectroscopic sample: SPEC\_Z vs DNF\_Z (left), SPEC\_Z vs EAzY\_Z (right) and DNF\_Z vs EAzY\_Z (bottom)}
   \label{fig:DNF_EAZY_spec}%
\end{figure*}

We analysed the photo-zs obtained using both methods. Firstly, we selected from Y3 Deep Fields catalogue those galaxies with spectroscopic redshift information, $mag(i)<28.0$, flux measurements in the eight filters. This sample contains $55,198$ galaxies and covers a large portion of the total sample as we can see in the right panel of Fig.~\ref{fig:deepfield}. Figure~\ref{fig:DNF_EAZY_spec} shows the scatter of the photo-zs determined by both methods versus the spectroscopic redshift (on the left DNF and on the right EAzY). The metrics obtained by DNF slightly outperforms those provided by EAzY, justified in part for the completeness of the training sample used in this test.  The bias and the $\sigma_{68}^{Norm}$ are $-0.0148$ and $0.0519$ for EAzY and
$0.0065$ and $0.0390$ for DNF respectively. Both methods give good photo-zs values for $z < 1$. However, for $z \ge 1$ EAzY shows a somewhat biased behaviour. The plot at the bottom shows the photo-z values of DNF (X-axis) vs EAzY (Y-axis). We can see the same bias appeared on right panel, that is EAzY with respect to SPEC\_Z. Therefore, this behaviour seems to come from the EAzY estimation. It may be due to the lack of Y-band data. The break is poorly constrained from $z \sim 1$ until the 4000$\textup{\r{A}}$ break starts to enter the J-band. The prior tends to favour a lower redshift and so the point estimates are pulled to lower redshift slightly.  This would be partially alleviated with the full EAzY PDFs.

\begin{figure*}
   \centering
   \includegraphics[width=0.49\textwidth]{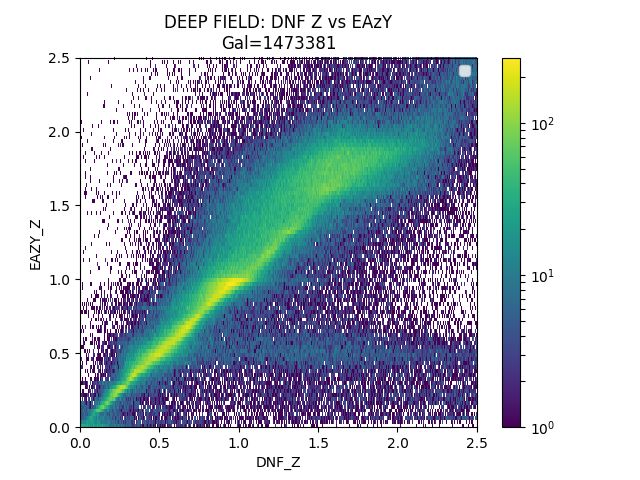}
   \includegraphics[width=0.49\textwidth]{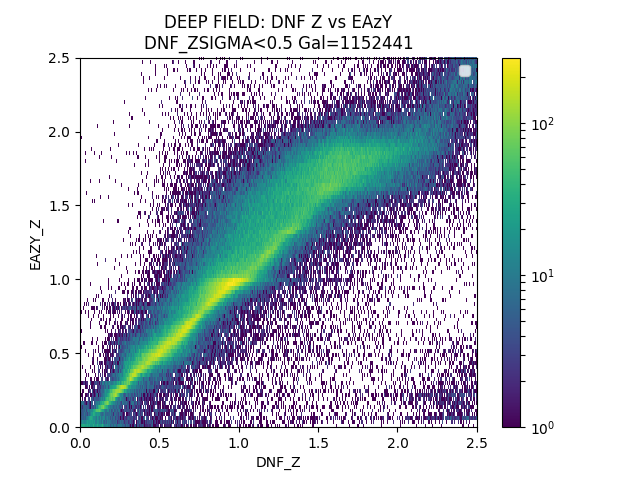}
   \caption{Y3 DES Deep Fields catalogue. Scatter plot DNF\_Z vs EAzY\_Z: for $mag<28$ (left) and for additional quality cut DNF\_ZSIGMA<0.5 (right) }
   \label{fig:DNFvsEAzY}
\end{figure*}

In Fig.~\ref{fig:DNFvsEAzY} (left), we compare the photo-z provided by both methods for the Y3 DES Deep Fields catalogue (right panel). We have focussed on galaxies with flux measurements in all the eight filters and $mag(i)<28.0$. It corresponds to a sample of 1,473,381 galaxies. For $z>1$ we can see a similar behaviour to that observed with spectroscopic redshifts (right panel of Fig.~\ref{fig:DNF_EAZY_spec}). Therefore, this behaviour seems to come from EAzY estimation. On the other hand, there is a cloud of points below the diagonal around $\mathrm{EAzY\_Z} \sim 0.5$ which extend along several values of $\mathrm{DNF\_Z}$. We can see in  Fig.~\ref{fig:DNFvsEAzY} (right) that the cloud can be removed by applying the quality cut $\mathrm{DNF\_ZSIGMA} <0.5$.  In general $\mathrm{DNF\_ZSIGMA}$ allow us to detect galaxies with large errors due to bad photometry, degeneracies or incompleteness.

Determining the best method to be applied to a scientific sample is non-trivial. \citet{2019NatAs...3..212S} points out that machine learning methods outperforms the template approaches when the training survey is sufficiently complete. However, template methods are more favourable when spectroscopic samples are limited. In the case of DNF and EAzY, the biggest differences appear for $z>1$, when the completeness of the training sample is poorer. Nevertheless, the photo-zs generated by DNF present better metrics than those of provided by EAzY. According to \citet{2019NatAs...3..212S}, template methods work best for high redshift because of the lack of photometric information to construct training samples for machine leaning methods. In the same sense, the templates are built on physical assumptions that may not be entirely correct or have incomplete coverage in certain areas.

\section{Conclusions} \label{sec:conclusion}
This study is an analysis of how the completeness and spectroscopic quality of the training sample affects the photometric redshift determination using the DNF algorithm. The conclusions are the following:

\begin{enumerate}

    \item We have emulated the problem of an incomplete training sample for DNF with the goal of measuring its effects and to take them into account on the photo-z performance. The principal component analysis provides a graphical method to assess completeness and $\mathrm{DNF\_ZSIGMA}$ turns out to be a reliable parameter to separate the set of galaxies computed with a complete training.
    
    \item We have analysed the possibility of substituting $z_{spec}$ by $\mathrm{DNF\_ZN}$ to assess $\Delta(z)$ in the scatter metrics of $\mathrm{DNF\_Z}$
    ($\mathrm{MAD}(\Delta z)$ and $\sigma_{68}^{Norm}$). The results show that $\mathrm{DNF\_ZN}$ provides an upper limit of the real values. Using this method the photo-z quality can be estimated when no spectroscopic information is available. 
    
    \item We determine the photo-zs of the Y3 DES Deep Fields catalogue using both, a semi-complete training sample with high and medium quality redshift spectroscopy and an incomplete training sample with the highest quality redshift spectroscopy. The obtained results are globally better for the semi-complete sample in spite of its slight relaxation in  quality. However, the photo-z improves for that sub-sample where the high-quality incomplete training covers its principal component analysis space. For faint magnitudes it seems better to use a training sample with medium quality spectroscopic redshift covering deeper magnitudes. This result advocates for training completeness at the expenses of sacrificing slightly the quality of the spectroscopic redshifts. The results also suggest that for specific studies focussed on brighter galaxies, we may be more interested in using only redshift of the highest possible quality in our training.
    
    \item We have compared the photometric redshift of Y3 DES Deep Fields catalogue determined with DNF and EAzY. Both methods show a similar behaviour up to $z \sim 1$. For $z>1$ DNF outperforms EAzY, which shows some bias towards higher redshift.

\end{enumerate}

\begin{acknowledgements}
      We acknowledges funding support from the Autonomous Community of Madrid through the project TEC2SPACE-CM (S2018/NMT-4291).

      Funding for the DES Projects has been provided by the U.S. Department of Energy, the U.S. National Science Foundation, the Ministry of Science and Education of Spain, 
the Science and Technology Facilities Council of the United Kingdom, the Higher Education Funding Council for England, the National Center for Supercomputing 
Applications at the University of Illinois at Urbana-Champaign, the Kavli Institute of Cosmological Physics at the University of Chicago, 
the Center for Cosmology and Astro-Particle Physics at the Ohio State University,
the Mitchell Institute for Fundamental Physics and Astronomy at Texas A\&M University, Financiadora de Estudos e Projetos, 
Funda{\c c}{\~a}o Carlos Chagas Filho de Amparo {\`a} Pesquisa do Estado do Rio de Janeiro, Conselho Nacional de Desenvolvimento Cient{\'i}fico e Tecnol{\'o}gico and 
the Minist{\'e}rio da Ci{\^e}ncia, Tecnologia e Inova{\c c}{\~a}o, the Deutsche Forschungsgemeinschaft and the Collaborating Institutions in the Dark Energy Survey. 

The Collaborating Institutions are Argonne National Laboratory, the University of California at Santa Cruz, the University of Cambridge, Centro de Investigaciones Energ{\'e}ticas, 
Medioambientales y Tecnol{\'o}gicas-Madrid, the University of Chicago, University College London, the DES-Brazil Consortium, the University of Edinburgh, 
the Eidgen{\"o}ssische Technische Hochschule (ETH) Z{\"u}rich, 
Fermi National Accelerator Laboratory, the University of Illinois at Urbana-Champaign, the Institut de Ci{\`e}ncies de l'Espai (IEEC/CSIC), 
the Institut de F{\'i}sica d'Altes Energies, Lawrence Berkeley National Laboratory, the Ludwig-Maximilians Universit{\"a}t M{\"u}nchen and the associated Excellence Cluster Universe, 
the University of Michigan, NSF's NOIRLab, the University of Nottingham, The Ohio State University, the University of Pennsylvania, the University of Portsmouth, 
SLAC National Accelerator Laboratory, Stanford University, the University of Sussex, Texas A\&M University, and the OzDES Membership Consortium.

Based in part on observations at Cerro Tololo Inter-American Observatory at NSF's NOIRLab (NOIRLab Prop. ID 2012B-0001; PI: J. Frieman), which is managed by the Association of Universities for Research in Astronomy (AURA) under a cooperative agreement with the National Science Foundation.

The DES data management system is supported by the National Science Foundation under Grant Numbers AST-1138766 and AST-1536171.
The DES participants from Spanish institutions are partially supported by MICINN under grants ESP2017-89838, PGC2018-094773, PGC2018-102021, SEV-2016-0588, SEV-2016-0597, and MDM-2015-0509, some of which include ERDF funds from the European Union. IFAE is partially funded by the CERCA program of the Generalitat de Catalunya.
Research leading to these results has received funding from the European Research
Council under the European Union's Seventh Framework Program (FP7/2007-2013) including ERC grant agreements 240672, 291329, and 306478.
We  acknowledge support from the Brazilian Instituto Nacional de Ci\^encia
e Tecnologia (INCT) do e-Universo (CNPq grant 465376/2014-2).

This manuscript has been authored by Fermi Research Alliance, LLC under Contract No. DE-AC02-07CH11359 with the U.S. Department of Energy, Office of Science, Office of High Energy Physics.
\end{acknowledgements}

%

\begin{thebibliography}{}
\bibitem[Abdalla et al.(2011)]{2011MNRAS.417.1891A} Abdalla, F.~B., Banerji, M., Lahav, O., et al.\ 2011, \mnras, 417, 1891. doi:10.1111/j.1365-2966.2011.19375.x

\bibitem[Abolfathi et al.(2018)]{2018ApJS..235...42A} Abolfathi, B., Aguado, D.~S., Aguilar, G., et al.\ 2018, \apjs, 235, 42. doi:10.3847/1538-4365/aa9e8a

\bibitem[Arnouts et al.(1999)]{1999MNRAS.310..540A} Arnouts, S., Cristiani, S., Moscardini, L., et al.\ 1999, \mnras, 310, 540. doi:10.1046/j.1365-8711.1999.02978.x

\bibitem[Bayliss et al.(2016)]{2016ApJS..227....3B} Bayliss, M.~B., Ruel, J., Stubbs, C.~W., et al.\ 2016, \apjs, 227, 3. doi:10.3847/0067-0049/227/1/3

\bibitem[Bazin et al.(2011)]{2011AA...534A..43B} Bazin, G., Ruhlmann-Kleider, V., Palanque-Delabrouille, N., et al.\ 2011, \aap, 534, A43. doi:10.1051/0004-6361/201116898

\bibitem[Beck et al.(2017)]{2017MNRAS.468.4323B} Beck, R., Lin, C.-A., Ishida, E.~E.~O., et al.\ 2017, \mnras, 468, 4323. doi:10.1093/mnras/stx687

\bibitem[Ben{\'\i}tez(2000)]{2000ApJ...536..571B} Ben{\'\i}tez, N.\ 2000, \apj, 536, 571. doi:10.1086/308947

\bibitem[Blake et al.(2016)]{2016MNRAS.462.4240B} Blake, C., Amon, A., Childress, M., et al.\ 2016, \mnras, 462, 4240. doi:10.1093/mnras/stw1990

\bibitem[Bolzonella et al.(2000)]{2000A&A...363..476B} Bolzonella, M., Miralles, J.-M., \& Pell{\'o}, R.\ 2000, \aap, 363, 476. doi:10.48550/arXiv.astro-ph/0003380

\bibitem[Bonnett et al.(2016)]{2016PhRvD..94d2005B} Bonnett, C., Troxel, M.~A., Hartley, W., et al.\ 2016, \prd, 94, 042005. doi:10.1103/PhysRevD.94.042005

\bibitem[Brammer et al.(2008)]{2008ApJ...686.1503B} Brammer, G.~B., van Dokkum, P.~G., \& Coppi, P.\ 2008, \apj, 686, 1503. doi:10.1086/591786

\bibitem[Brescia et al.(2021)]{2021FrASS...8...70B} Brescia, M., Cavuoti, S., Razim, O., et al.\ 2021, Frontiers in Astronomy and Space Sciences, 8, 70. doi:10.3389/fspas.2021.658229

\bibitem[Carrasco Kind \& Brunner(2013)]{2013MNRAS.432.1483C} Carrasco Kind, M. \& Brunner, R.~J.\ 2013, \mnras, 432, 1483. doi:10.1093/mnras/stt574

\bibitem[Castander et al.(2012)]{2012SPIE.8446E..6DC} Castander, F.~J., Ballester, O., Bauer, A., et al.\ 2012, \procspie, 8446, 84466D. doi:10.1117/12.926234

\bibitem[Cavuoti et al.(2012)]{2012A&A...546A..13C} Cavuoti, S., Brescia, M., Longo, G., et al.\ 2012, \aap, 546, A13. doi:10.1051/0004-6361/201219755

\bibitem[Cavuoti et al.(2017)]{2017MNRAS.466.2039C} Cavuoti, S., Tortora, C., Brescia, M., et al.\ 2017, \mnras, 466, 2039. doi:10.1093/mnras/stw3208

\bibitem[Childress et al.(2017)]{2017MNRAS.472..273C} Childress, M.~J., Lidman, C., Davis, T.~M., et al.\ 2017, \mnras, 472, 273. doi:10.1093/mnras/stx1872

\bibitem[Coil et al.(2011)]{2011ApJ...741....8C} Coil, A.~L., Blanton, M.~R., Burles, S.~M., et al.\ 2011, \apj, 741, 8. doi:10.1088/0004-637X/741/1/8

\bibitem[Colless et al.(2001)]{2001MNRAS.328.1039C} Colless, M., Dalton, G., Maddox, S., et al.\ 2001, \mnras, 328, 1039. doi:10.1046/j.1365-8711.2001.04902.x

\bibitem[Collister \& Lahav(2004)]{2004PASP..116..345C} Collister, A.~A. \& Lahav, O.\ 2004, \pasp, 116, 345. doi:10.1086/383254 Blanton, M.~R., et al.\ 2013, \apj, 767, 118. doi:10.1088/0004-637X/767/2/118

\bibitem[Cool et al.(2013)]{2013ApJ...767..118C} Cool, R.~J., Moustakas, J., Blanton, M.~R., et al.\ 2013, \apj, 767, 118. doi:10.1088/0004-637X/767/2/118

\bibitem[Cooper et al.(2012)]{2012MNRAS.425.2116C} Cooper, M.~C., Yan, R., Dickinson, M., et al.\ 2012, \mnras, 425, 2116. doi:10.1111/j.1365-2966.2012.21524.x

\bibitem[Davis et al.(2017)]{2017arXiv171002517D} Davis, C., Gatti, M., Vielzeuf, P., et al.\ 2017, arXiv:1710.02517. doi:10.48550/arXiv.1710.02517

\bibitem[Davis et al.(2003)]{2003SPIE.4834..161D} Davis, M., Faber, S.~M., Newman, J., et al.\ 2003, \procspie, 4834, 161. doi:10.1117/12.457897

\bibitem[De Vicente et al.(2016)]{2016MNRAS.459.3078D} De Vicente, J., S{\'a}nchez, E., \& Sevilla-Noarbe, I.\ 2016, \mnras, 459, 3078. doi:10.1093/mnras/stw857

\bibitem[DESI Collaboration et al.(2016)]{2016arXiv161100036D} DESI Collaboration, Aghamousa, A., Aguilar, J., et al.\ 2016, arXiv:1611.00036. doi:10.48550/arXiv.1611.00036

\bibitem[Driver et al.(2011)]{2011MNRAS.413..971D} Driver, S.~P., Hill, D.~T., Kelvin, L.~S., et al.\ 2011, \mnras, 413, 971. doi:10.1111/j.1365-2966.2010.18188.x

\bibitem[Euclid Collaboration et al.(2020)]{2020A&A...644A..31E} Euclid Collaboration, Desprez, G., Paltani, S., et al.\ 2020, \aap, 644, A31. doi:10.1051/0004-6361/202039403

\bibitem[Flaugher et al.(2015)]{2015AJ....150..150F} Flaugher, B., Diehl, H.~T., Honscheid, K., et al.\ 2015, \aj, 150, 150. doi:10.1088/0004-6256/150/5/150

\bibitem[Garilli et al.(2008)]{2008AA...486..683G} Garilli, B., Le F{\`e}vre, O., Guzzo, L., et al.\ 2008, \aap, 486, 683. doi:10.1051/0004-6361:20078878

\bibitem[Garilli et al.(2014)]{2014AA...562A..23G} Garilli, B., Guzzo, L., Scodeggio, M., et al.\ 2014, \aap, 562, A23. doi:10.1051/0004-6361/201322790

\bibitem[Geha et al.(2017)]{2017ApJ...847....4G} Geha, M., Wechsler, R.~H., Mao, Y.-Y., et al.\ 2017, \apj, 847, 4. doi:10.3847/1538-4357/aa8626

\bibitem[Gschwend et al.(2018)]{2018AC....25...58G} Gschwend, J., Rossel, A.~C., Ogando, R.~L.~C., et al.\ 2018, Astronomy and Computing, 25, 58. doi:10.1016/j.ascom.2018.08.008

\bibitem[Hartley et al.(2020)]{2020MNRAS.496.4769H} Hartley, W.~G., Chang, C., Samani, S., et al.\ 2020, \mnras, 496, 4769. doi:10.1093/mnras/staa1812

\bibitem[Hartley et al.(2022)]{2022MNRAS.509.3547H} Hartley, W.~G., Choi, A., Amon, A., et al.\ 2022, \mnras, 509, 3547. doi:10.1093/mnras/stab3055

\bibitem[Hildebrandt et al.(2010)]{2010A&A...523A..31H} Hildebrandt, H., Arnouts, S., Capak, P., et al.\ 2010, \aap, 523, A31. doi:10.1051/0004-6361/201014885

\bibitem[Ilbert et al.(2006)]{2006A&A...457..841I} Ilbert, O., Arnouts, S., McCracken, H.~J., et al.\ 2006, \aap, 457, 841. doi:10.1051/0004-6361:20065138

\bibitem[Jones et al.(2009)]{2009MNRAS.399..683J} Jones, D.~H., Read, M.~A., Saunders, W., et al.\ 2009, \mnras, 399, 683. doi:10.1111/j.1365-2966.2009.15338.x

\bibitem[Kaiser et al.(2010)]{2010SPIE.7733E..0EK} Kaiser, N., Burgett, W., Chambers, K., et al.\ 2010, \procspie, 7733, 77330E. doi:10.1117/12.859188

\bibitem[Le F{\`e}vre et al.(2004)]{2004AA...428.1043L} Le F{\`e}vre, O., Vettolani, G., Paltani, S., et al.\ 2004, \aap, 428, 1043. doi:10.1051/0004-6361:20048072

\bibitem[Le F{\`e}vre et al.(2005)]{2005A&A...439..845L} Le F{\`e}vre, O., Vettolani, G., Garilli, B., et al.\ 2005, \aap, 439, 845. doi:10.1051/0004-6361:20041960

\bibitem[Lidman et al.(2016)]{2016PASA...33....1L} Lidman, C., Ardila, F., Owers, M., et al.\ 2016, \pasa, 33, e001. doi:10.1017/pasa.2015.52

\bibitem[Lidman et al.(2013)]{2013PASA...30....1L} Lidman, C., Ruhlmann-Kleider, V., Sullivan, M., et al.\ 2013, \pasa, 30, e001. doi:10.1017/pasa.2012.001

\bibitem[Lilly et al.(2009)]{2009ApJS..184..218L} Lilly, S.~J., Le Brun, V., Maier, C., et al.\ 2009, \apjs, 184, 218. doi:10.1088/0067-0049/184/2/218

\bibitem[Lima et al.(2008)]{2008MNRAS.390..118L} Lima, M., Cunha, C.~E., Oyaizu, H., et al.\ 2008, \mnras, 390, 118. doi:10.1111/j.1365-2966.2008.13510.x

\bibitem[LSST Science Collaboration et al.(2009)]{2009arXiv0912.0201L} LSST Science Collaboration, Abell, P.~A., Allison, J., et al.\ 2009, arXiv:0912.0201. doi:10.48550/arXiv.0912.0201

\bibitem[Mao et al.(2012)]{2012MNRAS.426.3334M} Mao, M.~Y., Sharp, R., Norris, R.~P., et al.\ 2012, \mnras, 426, 3334. doi:10.1111/j.1365-2966.2012.21913.x

\bibitem[Masters et al.(2017)]{2017ApJ...841..111M} Masters, D.~C., Stern, D.~K., Cohen, J.~G., et al.\ 2017, \apj, 841, 111. doi:10.3847/1538-4357/aa6f08

\bibitem[Momcheva et al.(2016)]{2016ApJS..225...27M} Momcheva, I.~G., Brammer, G.~B., van Dokkum, P.~G., et al.\ 2016, \apjs, 225, 27. doi:10.3847/0067-0049/225/2/27

\bibitem[Muzzin et al.(2012)]{2012ApJ...746..188M} Muzzin, A., Wilson, G., Yee, H.~K.~C., et al.\ 2012, \apj, 746, 188. doi:10.1088/0004-637X/746/2/188

\bibitem[Nanayakkara et al.(2016)]{2016ApJ...828...21N} Nanayakkara, T., Glazebrook, K., Kacprzak, G.~G., et al.\ 2016, \apj, 828, 21. doi:10.3847/0004-637X/828/1/21

\bibitem[Nord et al.(2016)]{2016ApJ...827...51N} Nord, B., Buckley-Geer, E., Lin, H., et al.\ 2016, \apj, 827, 51. doi:10.3847/0004-637X/827/1/51

\bibitem[Parkinson et al.(2012)]{2012PhRvD..86j3518P} Parkinson, D., Riemer-S{\o}rensen, S., Blake, C., et al.\ 2012, \prd, 86, 103518. doi:10.1103/PhysRevD.86.103518

\bibitem[Rest et al.(2014)]{2014ApJ...795...44R} Rest, A., Scolnic, D., Foley, R.~J., et al.\ 2014, \apj, 795, 44. doi:10.1088/0004-637X/795/1/44

\bibitem[Sadeh et al.(2016)]{2016PASP..128j4502S} Sadeh, I., Abdalla, F.~B., \& Lahav, O.\ 2016, \pasp, 128, 104502. doi:10.1088/1538-3873/128/968/104502

\bibitem[Salvato et al.(2019)]{2019NatAs...3..212S} Salvato, M., Ilbert, O., \& Hoyle, B.\ 2019, Nature Astronomy, 3, 212. doi:10.1038/s41550-018-0478-0

\bibitem[Schmidt et al.(2020)]{2020MNRAS.499.1587S} Schmidt, S.~J., Malz, A.~I., Soo, J.~Y.~H., et al.\ 2020, \mnras, 499, 1587. doi:10.1093/mnras/staa2799

\bibitem[Scolnic et al.(2014)]{2014ApJ...795...45S} Scolnic, D., Rest, A., Riess, A., et al.\ 2014, \apj, 795, 45. doi:10.1088/0004-637X/795/1/45

\bibitem[Silverman et al.(2015)]{2015ApJS..220...12S} Silverman, J.~D., Kashino, D., Sanders, D., et al.\ 2015, \apjs, 220, 12. doi:10.1088/0067-0049/220/1/12

\bibitem[Stalin et al.(2010)]{2010MNRAS.401..294S} Stalin, C.~S., Petitjean, P., Srianand, R., et al.\ 2010, \mnras, 401, 294. doi:10.1111/j.1365-2966.2009.15636.x

\bibitem[Sullivan et al.(2011)]{2011yCat..74060782S} Sullivan, M., Conley, A., Howell, D.~A., et al.\ 2011, VizieR Online Data Catalog, J/MNRAS/406/782

\bibitem[S{\'a}nchez et al.(2014)]{2014MNRAS.445.1482S} S{\'a}nchez, C., Carrasco Kind, M., Lin, H., et al.\ 2014, \mnras, 445, 1482. doi:10.1093/mnras/stu1836

\bibitem[Tanaka et al.(2018)]{2018PASJ...70S...9T} Tanaka, M., Coupon, J., Hsieh, B.-C., et al.\ 2018, \pasj, 70, S9. doi:10.1093/pasj/psx077

\bibitem[Tasca et al.(2017)]{2017AA...600A.110T} Tasca, L.~A.~M., Le F{\`e}vre, O., Ribeiro, B., et al.\ 2017, \aap, 600, A110. doi:10.1051/0004-6361/201527963

\bibitem[Treu et al.(2015)]{2015ApJ...812..114T} Treu, T., Schmidt, K.~B., Brammer, G.~B., et al.\ 2015, \apj, 812, 114. doi:10.1088/0004-637X/812/2/114

\bibitem[York et al.(2000)]{2000AJ....120.1579Y} York, D.~G., Adelman, J., Anderson, J.~E., et al.\ 2000, \aj, 120, 1579. doi:10.1086/301513

\bibitem[Yuan et al.(2015)]{2015MNRAS.452.3047Y} Yuan, F., Lidman, C., Davis, T.~M., et al.\ 2015, \mnras, 452, 3047. doi:10.1093/mnras/stv1507



\end{thebibliography}
%

%

\begin{appendix} 
\section{Spectroscopic data}\label{sec:appendixA}
We have listed in Table \ref{tab:spec_data} the 34 spectroscopic samples compiled by \cite{2018AC....25...58G} to create the spectroscopic sample used in this work.

\begin{table*}[!ht]
    \centering
    \caption[]{Spectroscopic samples used in \cite{2018AC....25...58G}.}
    \label{tab:spec_data}
    \begin{tabular}{ll}
        \hline
        \textbf{Survey} & \textbf{Ref.} \\
        \hline
        PRIMUS & \cite{2011ApJ...741....8C, 2013ApJ...767..118C} \\
        SDSS DR14 & \cite{2018ApJS..235...42A} \\
        DES AAOmega & \cite{2015MNRAS.452.3047Y, 2017MNRAS.472..273C} \\
        VIPERS & \cite{2014AA...562A..23G} and \href{http://vipers.inaf.it/rel-pdr1.html}{http://vipers.inaf.it/rel-pdr1.html} \\
        WiggleZ & \cite{2012PhRvD..86j3518P}   and \href{http://wigglez.swin.edu.au/site/}{http://wigglez.swin.edu.au/site/} \\
        VVDS & \cite{2008AA...486..683G, 2004AA...428.1043L} \\ 
        zCOSMOS & \cite{2009ApJS..184..218L}  \\
        3D-HST & \cite{2016ApJS..225...27M} and \href{http://3dhst.research.yale.edu/Data.php}{http://3dhst.research.yale.edu/Data.php} \\
        DEEP2 & \cite{2003SPIE.4834..161D, 2017arXiv171002517D} and \href{http://deep.ps.uci.edu/DR4/home.html}{http://deep.ps.uci.edu/DR4/home.html}  \\            
        2dF & \cite{2001MNRAS.328.1039C} and \href{http://www.2dfgrs.net/}{http://www.2dfgrs.net/} \\
        GAMA & \cite{2011MNRAS.413..971D}   \\
        ACES & \cite{2012MNRAS.425.2116C} and \href{http://mur.ps.uci.edu/cooper/ACES/zcatalog.html}{http://mur.ps.uci.edu/cooper/ACES/zcatalog.html}  \\
        6dF & \cite{2009MNRAS.399..683J} and \href{http://www.6dfgs.net/}{http://www.6dfgs.net/} \\
        DES\ IMACS & \cite{2016ApJ...827...51N}  \\
        SAGA & \cite{2017ApJ...847....4G} and \href{http://sagasurvey.org/}{http://sagasurvey.org/}  \\     NOAO OzDES & \cite{2015MNRAS.452.3047Y, 2017MNRAS.472..273C}\\ 
        XXL AAOmega & \cite{2016PASA...33....1L} and  \href{http://cosmosdb.iasf-milano.inaf.it/XXL/}{http://cosmosdb.iasf-milano.inaf.it/XXL/}   \\
        SPT\ GMOS & \cite{2016ApJS..227....3B}\\
        UDS & \href{http://www.nottingham.ac.uk/astronomy/UDS/UDSz/}{http://www.nottingham.ac.uk/astronomy/UDS/UDSz/}  \\
        SNLS\ FORS & \cite{2011AA...534A..43B}   \\
        ATLAS & \cite{2012MNRAS.426.3334M}   \\            
        Pan-STARRS & \cite{2014ApJ...795...44R, 2014ApJ...795...45S, 2010SPIE.7733E..0EK}  \\
        C3R2 & \cite{2017ApJ...841..111M}   \\
        SpARCS & \cite{2012ApJ...746..188M}   \\   
        SNVETO & \href{http://www.ast.cam.ac.uk/~fo250/Research/SNveto/}{http://www.ast.cam.ac.uk/~fo250/Research/SNveto/}  \\
        FMOS-COSMOS & \cite{2015ApJS..220...12S} and \href{http://member.ipmu.jp/fmos-cosmos/FC\_catalogs.html}{http://member.ipmu.jp/fmos-cosmos/FC\_catalogs.html}   \\
        SNLS\ AAOmega & \cite{2013PASA...30....1L, 2015MNRAS.452.3047Y, 2017MNRAS.472..273C} and \\
        & \href{http://apm5.ast.cam.ac.uk/arc-bin/wdb/aat\_database/observation\_log/make}{http://apm5.ast.cam.ac.uk/arc-bin/wdb/aat\_database/observation\_log/make}  \\  
        CDB &  \cite{2011yCat..74060782S} \\
        VUDS & \cite{2017AA...600A.110T} and \href{http://cesam.lam.fr/vuds/DR1/}{http://cesam.lam.fr/vuds/DR1/} \\
        ZFIRE & \cite{2016ApJ...828...21N} and \href{http://zfire.swinburne.edu.au/data.html}{http://zfire.swinburne.edu.au/data.html}   \\          MOSFIRE & \href{http://mosdef.astro.berkeley.edu}{http://mosdef.astro.berkeley.edu}  \\
        2dFLenS & \cite{2016MNRAS.462.4240B} and \href{http://2dflens.swin.edu.au/}{http://2dflens.swin.edu.au/}{http://2dflens.swin.edu.au/}  \\
        GLASS & \cite{2015ApJ...812..114T} and \href{ https://archive.stsci.edu/prepds/glass/}{ https://archive.stsci.edu/prepds/glass/}   \\            
        XMM-LSS & \cite{2010MNRAS.401..294S}  \\
        \hline
    \end{tabular}
\end{table*}

\section{DNF\_ZSIGMA as an indicator of the quality of photo-z}\label{sec:appendixB}

DNF\_ZSIGMA is the indicator of the quality of each photo-z provided by DNF. These values are computed from the quadratic sum of the error due to the photometry plus the error due to the fit. In this Appendix, we  analyse the DNF\_ZSIGMA values. For this purpose, we have calculated the \textit{pull} defined as follow:

$$pull = \frac{z_{spec} - \mathrm{DNF\_Z}}{\mathrm{DNF\_ZSIGMA}},$$
where $z_{spec}$ is the spectroscopic redshfit and DNF\_Z the photometric redshift. 

Figure~\ref{fig:pull} compares the \textit{pull} distribution (blue) with a standard Gaussian distribution  with mean zero and unit width (orange line) for the values obtained from the complete sample. The \textit{pull} together with the central limit theorem allows us to analyse the possible dispersion and bias in the DNF\_ZSIGMA values comparing the \textit{pull} distribution with a standard Gaussian.  The results obtained from the \textit{pull} using the complete training sample fit to the Gaussian distribution. The \textit{pull} distribution is slightly narrower in the centre and with larger wings. These differences are showing that DNF\_ZSIGMA overestimates the errors for photo-zs with small errors and underestimates for large errors.

\begin{figure}
   \centering
    \includegraphics[width=0.49\textwidth]{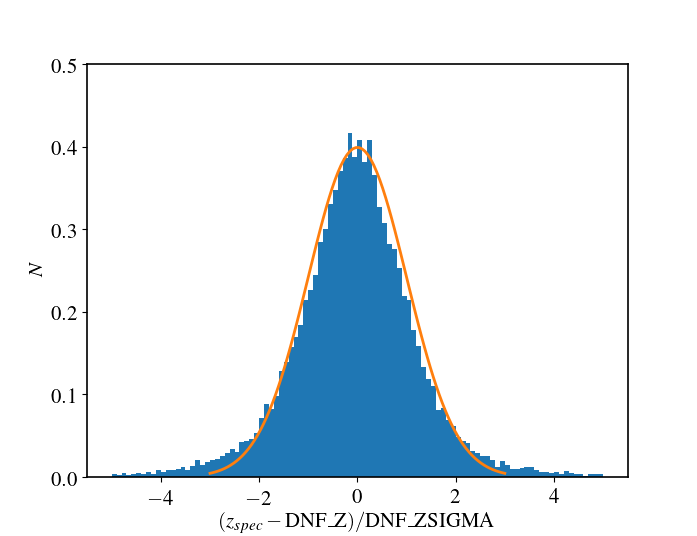} 
   \caption{Comparison of \textit{pull} distribution (blue) with a standard Gaussian distribution (orange line).}
   \label{fig:pull}
\end{figure}

\begin{figure}
   \centering
    \includegraphics[width=0.49\textwidth]{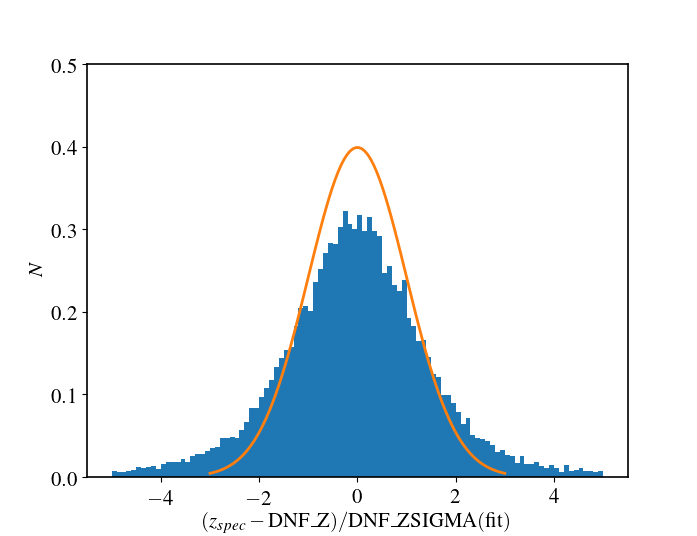} 
    \includegraphics[width=0.49\textwidth]{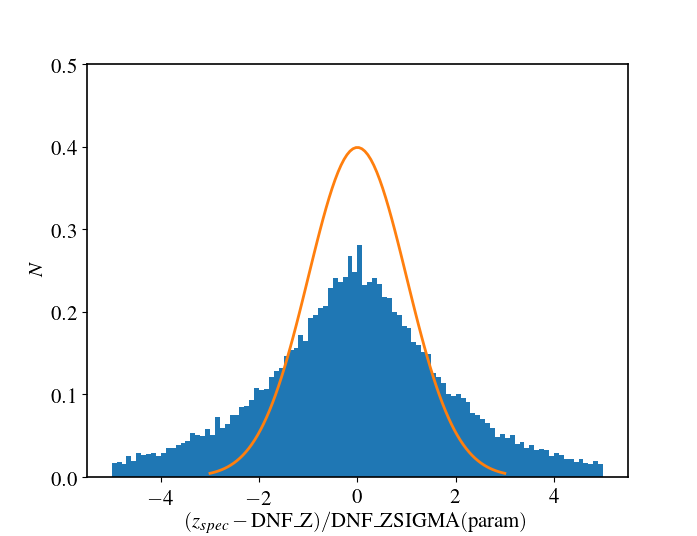}
   \caption{Comparison of \textit{pull} distribution (blue) with a standard Gaussian distribution (orange line) for the error due to the fit (upper panel) and to the photometry (bottom panel).}
   \label{fig:pull}
\end{figure}

\section{Effect of training sample size on Photometric Redshift}\label{sec:appendixB2}
\begin{figure}
   \centering
    \includegraphics[width=0.49\textwidth]{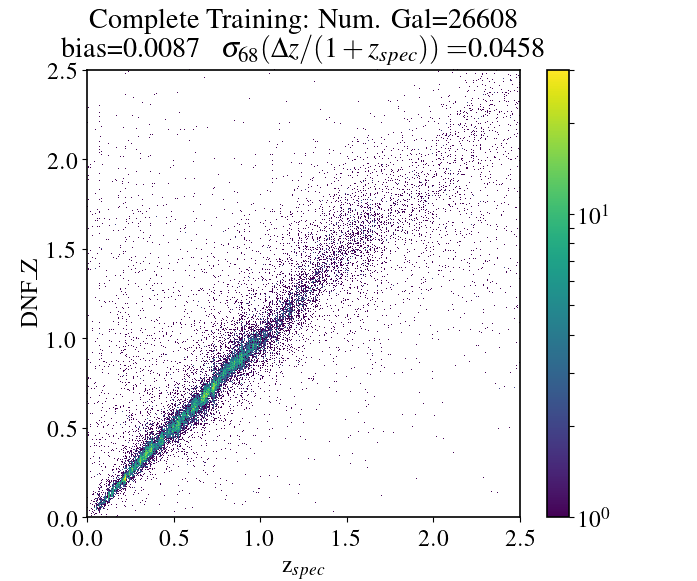} 
   \caption{Scatter plot of spectroscopic redshift $z_{spec}$ and the photo-z $\mathrm{DNF\_Z}$ for complete training of $5,336$ galaxies.}
   \label{fig:scatter_complete_5000}
\end{figure}

In addition to incompleteness, the number of galaxies in the training sample is also a factor that must be taken into account to determine the quality of the photometric redshift. We wanted to check what the results would be if the complete sample had the same number of galaxies as our incomplete sample, that is, $5,336$ galaxies. In this appendix, the Fig.~\ref{fig:scatter_complete_5000} shows the comparison between the spectroscopic redshift ($z_{spec}$) and the photometric redshift ($\mathrm{DNF\_Z}$) for a training sample that is complete but consists of $5,336$ galaxies. The number of galaxies that DNF has calculated photometric redshifts for with the same cuts defined in \ref{sec:incomplete_ass} is $26,608$ galaxies ($95.7\%$ of the sample). This value is very close to the case of the complete sample with $27,801$ galaxies ($96.8\%$) and considerably improves the result of the incomplete sample ($81.3\%$). On the other hand, the results found are intermediate values between the incomplete and complete cases for bias and $\sigma_{68}^{Norm}$.

\section{Quality metrics as a function of the redshift}\label{sec:appendixC}

\begin{figure}
   \centering
   \includegraphics[width=0.49\textwidth]{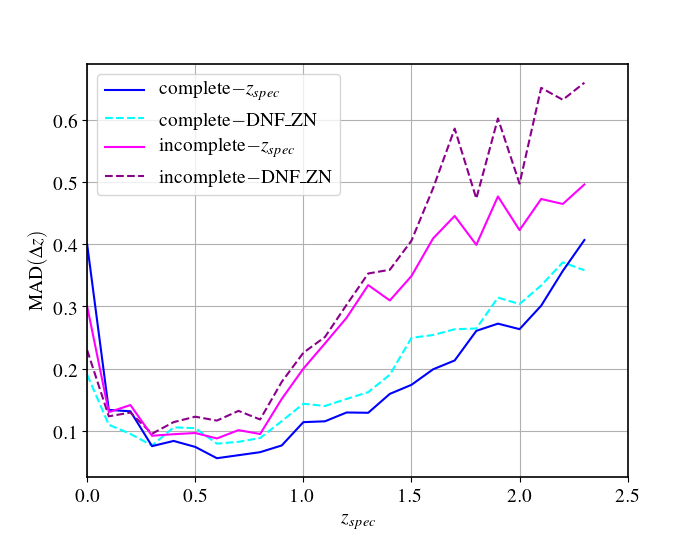}

   \caption{$\mathrm{MAD}(\Delta z)$ as a function of the $z_{spec}$ for the complete training (blue lines) and the incomplete training (magenta lines). The solid lines display the metrics calculated with $z_{spec}$ and the dashed lines replacing $z_{spec}$ by DNF\_ZN.}
   \label{fig:z_deltaz}%
\end{figure}

\begin{figure}
   \centering

   \includegraphics[width=0.49\textwidth]{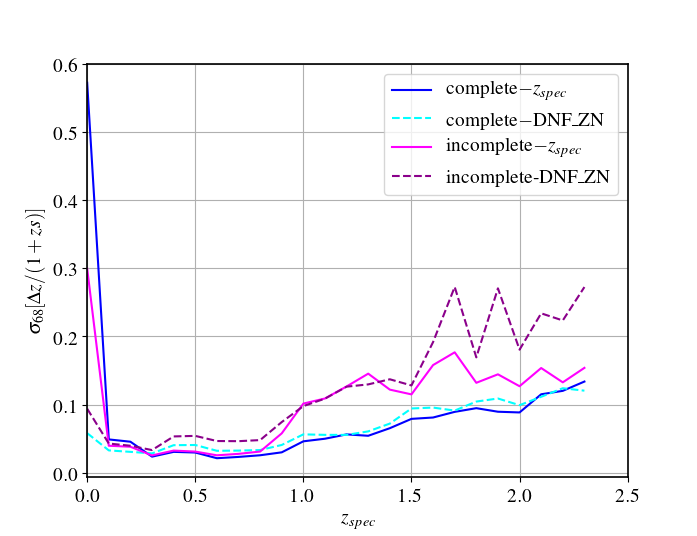} 

   \caption{$\sigma_{68}^{Norm}$ as a function of the $z_{spec}$ for the complete training (blue lines) and the incomplete training (magenta lines). The solid lines display the metrics calculated with $z_{spec}$ and the dashed lines replacing $z_{spec}$ by DNF\_ZN.}
   \label{fig:z_sigma68_appendix}%
\end{figure}

We studied the behaviour of the photo-zs estimation as a function of the spectroscopic redshift for the complete and incomplete spectroscopic training samples defined in Sect. \ref{sec:ref_effect_incompleteness}. The Fig.~\ref{fig:z_deltaz} and \ref{fig:z_sigma68_appendix} show the behaviour of the absolute median deviation  and the $\sigma_{68}^{Norm}$ as a function of $z_{spec}$ for the complete training sample (blue lines) and incomplete training sample (magenta lines). We have also calculated the mean absolute deviation and the $\sigma_{68}^{Norm}$ replacing $z_{spec}$ by $\mathrm{DNF\_ZN}$ (dashed lines). As in Fig.~\ref{fig:mag_deltaz} of Sect. \ref{sec:photo-z_estimation_perf}, in both plots the behaviour of the mean absolute deviation and the $\sigma_{68}^{Norm}$ can be considered a good approximation to the real value which changes depending on the training sample. The high errors that can be observed for $z_{spec}$ close to zero are due to stars wrongly classified in the validation sample. 

\begin{figure}
   \centering
    \includegraphics[width=0.49\textwidth]{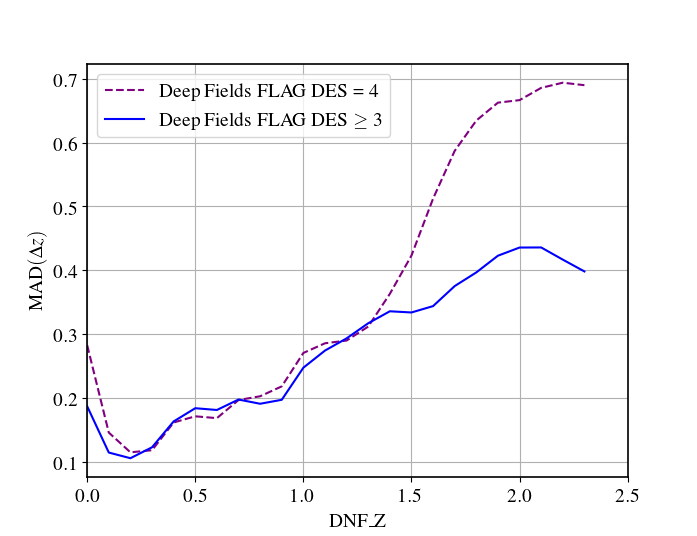} 

  \caption{The precision of the photo-z estimates defined by the absolute median deviation as a function of the $z_{spec}$ calculated by $\mathrm{DNF\_Z}$ and $\mathrm{DNF\_ZN}$ determined by training sample with only galaxies with FLAG\_DES $=$ 4 (incomplete training sample) and training sample with galaxies with FLAG\_DES$<=$3 (semi-complete training sample), in purple and blue, respectivaly.}
   \label{fig:deepfield_mad_Z}%
\end{figure}

\begin{figure}
   \centering
   \includegraphics[width=0.49\textwidth]{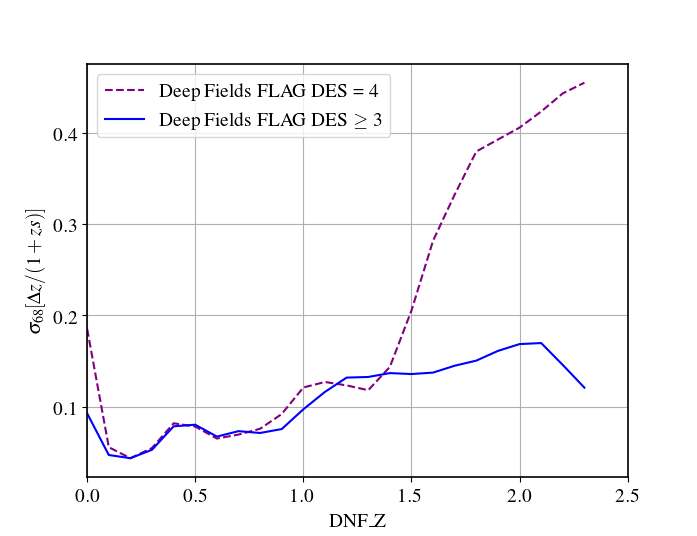}

  \caption{The precision of the photo-z estimates defined by the $\sigma_{68}^{Norm}$ as a function of the $z_{spec}$ calculated by $\mathrm{DNF\_Z}$ and $\mathrm{DNF\_ZN}$ determined by training sample with only galaxies with FLAG\_DES $=$ 4 (incomplete training sample) and training sample with galaxies with FLAG\_DES$<=$3 (semi-complete training sample), in purple and blue, respectivaly.}
   \label{fig:deepfield_metrics_Z}%
\end{figure}

In addition, we studied the behaviour of the photo-zs estimation as a function of the redshift for the galaxies of Y3 Deep Field catalogue using the incomplete and semi-incomplete training sample defined in Sect. \ref{sec:photoz_deepFields}. In this case, as we lack information
on the spectroscopic redshift, we have replaced $z_{spec}$ by DNF\_Z. The results of Fig.~\ref{fig:deepfield_mad_Z} and \ref{fig:deepfield_metrics_Z} show that $\mathrm{MAD}(\Delta z)$ and $\sigma_{68}^{Norm}$ get worse for higher redshift. Both training samples have similar results for $z < 1.4$. After this value, semi-incomplete training sample works better than incomplete training sample. 

\end{appendix}

\end{document}